\documentclass[aps,prd,amssymb,amsmath,showpacs]{revtex4}
\usepackage{graphicx}

\begin{document}

\title{
Real time approach to tunneling in open quantum systems:
decoherence and anomalous diffusion }
\author{Esteban Calzetta}
\affiliation{Departmento de F\'{\i}sica, Facultad de Ciencias
Exactas y Naturales, Universidad de Buenos Aires, Ciudad
Universitaria, Pabell\'on I, 1428 Buenos Aires, Argentina}
\author{Enric Verdaguer}
\affiliation{Departament de F\'{\i}sica Fonamental and CER en
Astrof\'\i sica, F\'\i sica de Part\'\i cules i Cosmologia,
Universitat de Barcelona, Av.~Diagonal 647, 08028 Barcelona,
Spain}



\begin{abstract}
Macroscopic quantum tunneling is described using the master
equation for the reduced Wigner function of an open quantum system
at zero temperature. Our model consists of a particle trapped in a
cubic potential interacting with an environment characterized by
dissipative and normal and anomalous diffusion coefficients. A
representation based on the energy eigenfunctions of the isolated
system, \textit{i.\;e.} the system uncoupled to the environment,
is used to write the reduced Wigner function, and the master
equation becomes simpler in that representation. The energy
eigenfunctions computed in a WKB approximation incorporate the
tunneling effect of the isolated system and the effect of the
environment is described by an equation that it is in many ways
similar to a Fokker-Planck equation. Decoherence is easily
identified from the master equation and we find that when the
decoherence time is much shorter than the tunneling time the
master equation can be approximated by a Kramers like equation
describing thermal activation due to the zero point fluctuations
of the quantum environment. The effect of anomalous diffusion can
be dealt with perturbatively and its overall effect is to inhibit
tunneling.

\end{abstract}

\maketitle


\section{Introduction}

\label{sec1}

The observation of quantum tunneling effects is now possible in
some macroscopic quantum variables such as the flux quantum
transitions in a superconducting quantum interference device, or
the decay of a zero-voltage state in a current-biased Josephson
junction \cite{DevMarCla85,MarDevCla87,CleMarCla88,WalEtAl03}.
Macroscopic quantum systems pertains to the boundary between
quantum and classical physics. These systems are modelled by open
quantum systems which are characterized by a distinguished
subsystem, the ``system'' for short, interacting with an
environment. The environment acts as a source of dissipation and
noise for the system and produces quantum decoherence which
generally inhibits tunneling \cite{CalLeg81,CalLeg83b}. There is a
vast literature on this subject, see Refs. \cite
{LegEtAl87,HanTalBor90,Wei93} for comprehensive reviews.

Most work on macroscopic quantum tunneling is based on imaginary
time formalisms such as the Euclidean functional techniques which
have been introduced in the classical field of noise-activated
escape from a metaestable state \cite{Lan67}, or the instanton
approach introduced for quantum mechanical tunneling or for vacuum
decay in field theory \cite
{VolKobOku75,Col77,CalCol77,ColGlaMar78,ColDeL80,Col85}. These
techniques are specially suited for equilibrium or near
equilibrium situations, but are difficult to generalize to non
equilibrium situations.

There are theoretical and practical reasons for a formalism of non
equilibrium macroscopic quantum tunneling. On the theoretical side
dissipation as well as decoherence are only truly understood in a
dynamical real time formalism. In the classical context thermal
activation from metaestable states is well understood since
Kramers \cite{Kra40} in terms of the dynamical Fokker-Planck
transport equation, where the roles of dissipation and noise and
their inter-relations are known. On the other hand, an open
quantum system may be described by a dynamical equation for the
reduced density matrix, the so-called master equation, or the
equivalent equation for the reduced Wigner function which has many
similarities to the Fokker-Planck equation. However, at present no
compelling derivation of the tunneling rate is available in this
dynamical framework, that might be compared to the instanton
approach for equilibrium systems. Consequently, the effect of
dissipation, noise and decoherence on tunneling and their
inter-connections is not yet fully understood. On the practical
side out of equilibrium macroscopic quantum tunneling is becoming
necessary to understand arrays of Josephson junctions, or
time-dependent traps for cold atoms which are proposed for storing
quantum information in future quantum computers \cite
{ShnSchHer97,MooEtAl99,CleGel04,Mon02}, or to understand first
order phase transitions in cosmology \cite{Kib80,RivLomMaz02}.

In recent years we have considered different scenarios in which
metaestable quantum systems are described by the master equation
for the reduced Wigner function. By using techniques similar to
those used for thermal activation processes on metastable states
\cite{Kra40,Lan69} it was possible to compute the effect of the
environment on the quantum decay probability. This was used in
some semiclassical cosmological scenarios for noise induced
inflation \cite {CalVer99} due to the back reaction of the
inflaton field, in the context of stochastic semiclassical gravity
\cite {CalHu94,CamVer96,CalCamVer97,MarVer99a,MarVer99b}; see
Refs. \cite {HuVer03,HuVer04} for reviews on this subject. It was
also used for bubble nucleation in quantum field theory, where the
system was described by the homogeneous mode of the field of
bubble size and the environment was played by the inhomogeneous
modes of the field \cite{CalRouVer01a,CalRouVer02}, and on some
simple open quantum systems coupled linearly to a continuum of
harmonic oscillators at zero temperature \cite{ArtEtAl03}. But in
all these problems only the contribution to tunneling due to
activation was considered. One of the purposes of this paper is to
clarify when this approximation is valid.

In this paper we propose a formulation of macroscopic quantum
tunneling using the master equation for the reduced Wigner
function in which both the pure quantum tunneling effect and the
environment are taken into account. This is possible by the
introduction of a representation of the reduced Wigner function
based on the energy eigenfunctions of the isolated system,
\textit{i.\;e.} the system not coupled to the environment. This
representation is useful in a way somewhat analogous to the way
the energy representation is useful in the Schr\"{o}dinger
equation. The key to this result is that quantum tunneling is
already encoded in the energy eigenfunctions, which we can compute
in a WKB approximation. It is quite remarkable that in this
representation the master equation can be solved analytically
under certain approximations.

In order to have a working model in a form as simple as possible,
but that captures the main physics of the problem, we use constant
dissipation and normal and anomalous diffusion coefficients to
describe the effect of the environment. These coefficients can be
deduced from microscopic physics: they take constant values when
the environment is made by an Ohmic distribution of harmonic
oscillators weakly coupled in thermal equilibrium at high
temperature; but at zero temperature they are time dependent
\cite{HuPazZha92,ArtEtAl03,LomVil05}. Thus the model studied here
may be seen as a toy model at low temperature, generally valid at
long time scales only.

Master equations play also an important role in elucidating the
emergence of classicality in open quantum systems as a result of
their interaction with an environment. In fact, as the master
equation gives the quantum evolution of initial states, defined by
the reduced Wigner function at some initial time, it has been of
great help to study decoherence. In particular, it has been used
to clarify the way in which the environment selects a small set of
states of the system which are relatively stable by this
interaction, the so-called pointer states, whereas the coherent
superposition of the remaining states are rapidly destroyed by
decoherence \cite{Zur91,PazHabZur93,ZurPaz94,PazZur99,PazZur01}.
Using large scale numerical simulations the effect of the
interaction with the environment on coherent tunneling has been
analyzed in the framework of an open quantum system that is
classically chaotic: a harmonically driven quartic double well
\cite{MonPaz00,MonPaz01}. More recently \cite{AntEtAl05} tunneling
in a simple double well potential has been numerically simulated
using the master equation at high temperature as well as at zero
temperature. It is found that at zero temperature tunneling is
inhibited by the environment that produces decoherence
nevertheless at large time scales tunneling is still possible by
an activation-like process due to the zero point fluctuations of
the quantum environment. In the model we consider in this paper,
which involves a cubic potential, we find a very similar behavior
when the decoherence time is very short.

This paper is organized as follows. In the next two sections
\ref{sec2} and \ref{sec3} we review the theory of tunneling in
closed systems and introduce the energy representation for Wigner
functions. This extended review is necessary both to establish our
conventions and to recall specific results which are central to
the main argument. In section \ref{oqs} we introduce the
environment and write the master equation for the reduced Wigner
function of the open quantum system, discuss the different
dissipation and diffusion coefficients and derive the master
equation in the energy representation. In section \ref{ab} this
master equation is explicitly computed and the different time
scales of the problem are discussed. In section \ref{tunnel} under
the assumption of strong decoherence we analytically solve the
master equation, which is reduced to a Kramers-like equation, and
then perturbatively compute the effect of anomalous diffusion on
tunneling. Finally, in Section \ref{conclusions} we briefly
summarize our results. In the Appendices we provide additional
technical details.

\section{Tunneling in quantum mechanics}

\label{sec2}

In this section we review the WKB method to tunneling in quantum mechanics.
The energy eigenfunctions in the WKB approximation we obtain will play an
important role in the energy representation of the Wigner function that will
be introduced latter.

\subsection{The system}

\begin{figure}[tbp]
\includegraphics[height=8cm]{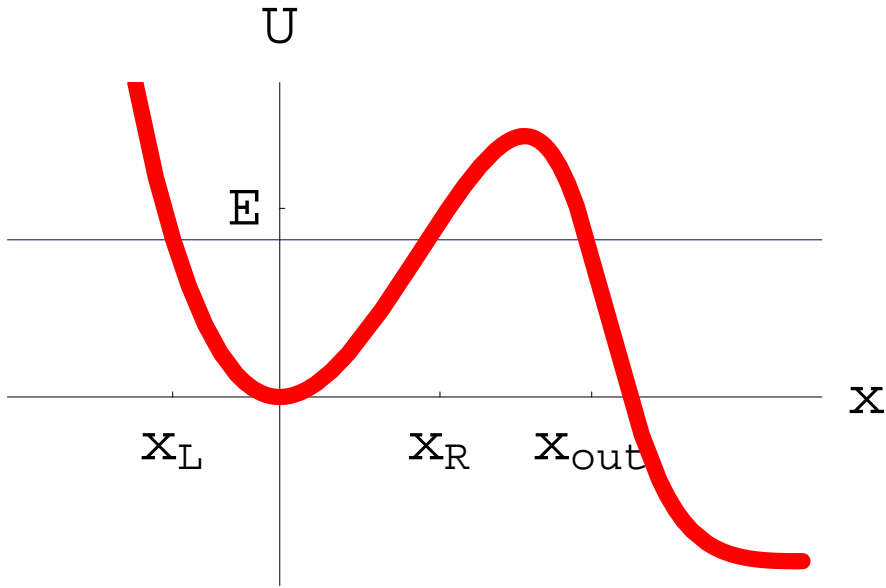}
\caption{A schematic plot of the potential. For an energy E below the
barrier there will be three classical turning points, also shown.}
\end{figure}

We begin with the simple closed quantum mechanical system formed by a
particle of mass $M$ in one dimension described by a Hamiltonian
\begin{equation}
H=\frac{p^{2}}{2M}+U\left( x\right) ,  \label{hamilton}
\end{equation}
with a potential $U$ given by
\begin{equation}
U\left( x\right) =\frac{1}{2}M\Omega _{0}^{2}x^{2}-\frac{\lambda }{6}x^{3},
\label{potential}
\end{equation}
for small values of the coordinate $x$. This is a fairly general potential
for a tunneling system, it is the basic element in the dashboard potential,
which is a very good model for a flux trapped in a superconducting quantum
interference device (SQUID), or a single Josephson junction biased by a
fixed external current \cite{CalLeg83b,MarDevCla87,Wei93,Tin96}. For
technical reasons, it is convenient to assume that for large $x$ the
potential flattens out and takes the value $U\left( x\right) =-U_{\infty },$
both negative and constant. The tunneling process ought to be independent of
the form of the potential this far away from the potential barrier. We
present a sketch of this potential in Fig. 1.

There is one classically stable point at $x=0$, and one unstable point $%
x=x_{s}=2M\Omega _{0}^{2}/\lambda ,$ \ corresponding to an energy $%
\varepsilon _{s}=2M^{3}\Omega _{0}^{6}/(3\lambda ^{2})$. The curvature of
the potential is $U^{\prime \prime }(0)=M\Omega _{0}^{2}$ at $x=0,$ and $%
-U^{\prime \prime }(0)$ \ at $x_{s}.$ The other point at which $U\left(
x\right) =0$ is $x=x_{exit}=(3/2)x_{s}.$ For $x\gg x_{exit}$ the potential
flattens out and is constant.

\subsection{The WKB approximation}

\label{wkb}

If we assume that the particle is trapped in the potential well, that is in
its false ground state or false vacuum, the tunneling probability can be
computed in this simple problem in many ways. One of the most efficient is
the instanton method which reduces to the computation of the ``bounce
solution''. The most attractive aspect of this computation is that it can be
easily extended to field theory where the tunneling probability is then
interpreted as the probability per unit time and volume to nucleate a bubble
of the true vacuum phase. The rate for quantum tunneling is $\Gamma
_{closed}=(\Omega _{0}/2\pi )a_{q}\exp (-S_{B}/\hbar )$, where $S_{B}$ is
the action for the ``bounce'' (or instanton), namely the solution to the
classical equations of motion which interpolates between $x=0$ and $%
x=x_{exit}$ in imaginary time
$S_{B}=2\int_{0}^{x_{exit}}dx\;\sqrt{2MU\left( x\right) },$ and
the prefactor $a_{q}=(120\pi S_{B}/\hbar )^{1/2}$. Our expression
for the potential is so simple that the above integral can be
computed explicitly: $S_{B}/\hbar =18\varepsilon
_{s}/(5\varepsilon _{0})$, where $\varepsilon
_{0}=\frac{1}{2}\hbar \Omega _{0}$ is the zero point energy of a
harmonic oscillator with frequency $\Omega _{0}.$

Here, however, we will concentrate on a real time approach by expanding the
false vacuum state as a linear combination of true eigenstates of the
Hamiltonian. To the required accuracy, it is enough to work with the WKB
approximations to the true eigenfunctions; see for instance Refs. \cite
{LanLif77,GalPas90}. The instanton method reviewed in the previous paragraph
can, in fact, be easily justified by this semiclassical approximation. Here
we explain in some detail this standard procedure to obtain the
eigenfunctions by matching the WKB solutions in the different regions of the
potential. These solutions will play a crucial role in the energy
representation for the Wigner functions to be introduced latter.

Let $0<E<\varepsilon _{s}$ be the energy of the particle in the potential
well, and $\psi _{E}$ the corresponding eigenfunction. The Schr\"{o}dinger
equation is
\begin{equation}
\frac{-\hbar ^{2}}{2M}\frac{\partial ^{2}}{\partial x^{2}}\psi _{E}+U\left(
x\right) \psi _{E}=E\psi _{E}.  \label{wkb1}
\end{equation}
Let us define
\begin{equation}
p\left( x\right) =\sqrt{2M\left| U\left( x\right) -E\right| },  \label{wkb2}
\end{equation}
and the integral $S(x,y)$ (note the order in the integration limits)
\begin{equation}
S\left( x,y\right) =\int_{y}^{x}dx^{\prime }\;p\left( x^{\prime }\right) .
\label{wkb3}
\end{equation}

The WKB solutions are obtained from these elements. We have to match the WKB
solutions in the different regions across the potential function. The
details of this calculation are given in Appendix A. The WKB solution $\psi
_{E}$ for energies in the range $0<E<\varepsilon _{s}$ is given by Eq. (\ref
{wkb10}), where $x_{L}<x_{R}<x_{out}$ are the three classical turning points
for the cubic potential (\ref{potential}); see Fig.~1. The normalization
constant $K_{E}$ in Eq. (\ref{wkb10}) is obtained by imposing the continuous
normalization of the eigenfunctions given in Eq. (\ref{norm1}) and it is
given in Eq. (\ref{norm6}). Of particular relevance is the value of the
eigenfunction $\psi _{E}(x)$ at values $x\gg x_{out}$. This gives the main
contribution to the continuous normalization integral. The value of the
eigenfunction at $x\gg x_{out}$, as computed in Appendix A, is
\begin{equation}
\psi _{E}(x)\sim \sqrt{\frac{2M}{\hbar \pi p_{\infty }}}\sin \left( \frac{%
p_{\infty }x}{\hbar }+\delta _{E}\right) ,  \label{norm8}
\end{equation}
where the phase $\delta _{E}$ is introduced in Eqs. (\ref{norm7}) and $%
p_{\infty }(E)$ is defined by Eq. (\ref{wkb2}) when $x\gg x_{out}$; see also
Eq. (\ref{norm2}).

We are interested in the details of the eigenfunctions near the false vacuum
state, since we will be dealing with tunneling from vacuum. Thus, in the
remaining of this section we give explicitly the values of the normalization
constant $K_{E}$ and the phase shifts $\delta _{E}$ near this vacuum state.
Therefore let us impose the Bohr-Sommerfeld quantization condition (\ref
{wkb11}) and let $E_{0}$ be the corresponding lowest energy, that is, $n=0$
in Eq. (\ref{wkb11}). As we will see in the next subsection this defines the
false vacuum energy. Expanding the integral in Eq. (\ref{wkb3}) around $E_{0}
$ we find that close to the lowest energy value
\begin{equation}
S\left( x_{R},x_{L}\right) \sim \frac{\pi \hbar }{2}-\tau \left(
E-E_{0}\right) ,  \label{norm11}
\end{equation}
where $\tau $ is defined by
\begin{equation}
\tau =\int_{x_{L}}^{x_{R}}dx\;\sqrt{\frac{2M}{U\left( x\right) -E_{0}}}.
\label{norm12}
\end{equation}
Thus $\cos ^{2}\left( S\left( x_{R},x_{L}\right) /\hbar \right) \sim (\tau
^{2}/\hbar ^{2})\left( E-E_{0}\right) ^{2}$, and evaluating the right hand
side of (\ref{norm10}) at $E_{0},$ we conclude that $K_{E}^{2}$ has poles at
the complex energies
\begin{equation}
E_{\pm }=E_{0}\pm i\varepsilon ,\qquad \varepsilon \equiv \frac{\hbar }{%
4\tau }e^{-2S_{0}\left( x_{out},x_{R}\right) /\hbar },
\label{norm13}
\end{equation}
which is in agreement with the standard result \cite{GalPas90}. To simplify
the notation let us call $S_{0}=S_{0}\left( x_{out},x_{R}\right) $ and $%
f_{0}=f\left( E_{0}\right) +\pi /4$, then we have from Eqs. (\ref{norm5a})
and (\ref{norm5b}) that the functions $A(E)$ and $B(E)$ for $E$ near $E_{0}$
are: $A\left( E\right) =(\tau /\hbar )\exp (S_{0}/\hbar )[F_{-}(E)+F_{+}(E)]$
and $B\left( E\right) =(-i\tau /\hbar )\exp (S_{0}/\hbar )[F_{-}(E)-F_{+}(E)]
$, where $F_{-}(E)=\exp (if_{0}/\hbar )\left( E-E_{-}\right) $ and $%
F_{+}(E)=\exp (-if_{0}/\hbar )\left( E-E_{+}\right) $. Notice that neither $A
$ nor $B$ vanish at $E_{\pm }$. Finally from Eq. (\ref{norm6}) we can write
the normalization constant near the false vacuum energy, as
\begin{equation}
K_{E}^{2}=\frac{M}{\pi \hbar \tau }\frac{\varepsilon }{\left( E-E_{0}\right)
^{2}+\varepsilon ^{2}}=\frac{4M\varepsilon ^{2}}{\pi \hbar ^{2}}\frac{%
e^{2S_{0}/\hbar }}{\left( E-E_{-}\right) \left( E-E_{+}\right) },
\label{norm15a}
\end{equation}
and from Eqs. (\ref{norm7}) the phase shifts are
\begin{equation}
e^{i\delta _{E}}=2\sqrt{\frac{\varepsilon ^{2}e^{2S_{0}/\hbar }}{\left(
E-E_{0}\right) ^{2}+\varepsilon ^{2}}}\left( A+iB\right) =e^{if_{0}/\hbar }%
\sqrt{\frac{E-E_{-}}{E-E_{+}}}.  \label{norm15b}
\end{equation}

Equations (\ref{norm8}), (\ref{norm15a}) and (\ref{norm15b}) are the main
results of this section. We notice, in particular, the poles of the norm and
the phase shifts at $E_\pm$ near the false vacuum energy. The strong
dependence on the energy of these functions near the false ground energy
will play an important role in the next sections. We will need, in particular, the
phase shifts derivatives which are given by:
\begin{equation}
\frac{\partial \delta_E }{\partial E}=\frac{-i}{2}\left( \frac{1}{E-E_{-}}-%
\frac{1}{E-E_{+}}\right) .  \label{phase4}
\end{equation}

\subsection{The false vacuum}

\label{vacu}

Before we start with the computation of the tunneling rate we have to define
what we mean by the decaying state, all the wave functions we considered so
far are true stationary states and, obviously, show no decay whatsoever. We
need to confine initially the particle into the potential well in its lowest
energy. To this end, we introduce an auxiliary potential $U_{\mathrm{aux}}$
which agrees with $U$ up to $x_{s}$ (where the true potential reaches its
maximum value) and increases thereafter. We may assume that the growth of $%
U_{\mathrm{aux}}$ is as fast as necessary to justify the approximations
below; the tunneling rate is insensitive to the details of $U_{\mathrm{aux}}$
beyond $x_{s}.$ Thus, we define the decaying state $\psi _{0}$ as the ground
state of a particle confined by $U_{\mathrm{aux}}$ \cite{Mig77}.

It is obvious from the form of the WKB solutions that $\psi _{0}$ agrees
with $\psi _{E_{0}}$ up to $x_{s}$, \textit{i.\ e.} $\psi _{0}(x)=\psi
_{E_{0}}(x)$ for $x\leq x_{s}$, where $E_{0}$ is the Bohr-Sommerfeld ground
state energy for the auxiliary potential $U_{\mathrm{aux}}$, which
corresponds to $n=0$ in the condition (\ref{wkb11}). Beyond $x_{s}$, $\psi
_{0}$ will decay rapidly to zero, unlike $\psi _{E_{0}}.$ Like any other
wave function, $\psi _{0}$ admits a development in the complete base of
energy eigenfunctions $\psi _{E}$, as
\begin{equation}
\psi _{0}\left( x\right) =\int dE\;C_{E}\psi _{E}\left( x\right) ,
\label{vacu1a}
\end{equation}
where due to our normalization the Fourier coefficients are given by
\begin{equation}
C_{E}=\int dx\;\psi _{E}\left( x\right) \psi _{0}\left( x\right) .
\label{vacu1b}
\end{equation}

To find these coefficients, we observe that $\psi _{0}\left( x\right) $ is a
solution to the Schr\"{o}dinger equation with the auxiliary potential $U_{%
\mathrm{aux}}$
\begin{equation}
\frac{-\hbar ^{2}}{2M}\frac{\partial ^{2}}{\partial x^{2}}\psi _{0}+U_{%
\mathrm{aux}}\left( x\right) \psi _{0}=E_{0}\psi _{0}.  \label{vacu2}
\end{equation}
Let us add to both sides of this equation the term $[U\left( x\right) -U_{%
\mathrm{aux}}\left( x\right) ]\psi _{0}$ and then multiply both sides by $%
\psi _{E}\left( x\right) $ and integrate to obtain
\begin{equation}
\left( E-E_{0}\right) C_{E}=-\int_{x_{s}}^{\infty }dx\;\psi _{E}\left(
x\right) \left[ U_{\mathrm{aux}}\left( x\right) -U\left( x\right) \right]
\psi _{0}\left( x\right) .  \label{new1}
\end{equation}
An important consideration is that $\psi _{0}\left( x\right) $ is a smooth
function (as opposed to a distribution), and, unlike $\psi _{E_{0}}$ it is
normalizable, so $C_{E}$ must also be smooth. This means that it is
allowable to assume $E\neq E_{0}$ in Eq. (\ref{new1}); $C_{E_{0}}$ can then
be found by analytical continuation. To estimate the right hand side of Eq. (%
\ref{new1}), let us introduce; cf. Eq. (\ref{wkb2}),
\begin{equation}
p_{\mathrm{aux}}\left( x\right) =\sqrt{2M\left| U_{\mathrm{aux}}\left(
x\right) -E_{0}\right| }.
\end{equation}
To the right of $x_{s}$ we may use the WKB approximation with the decaying
solution into the forbidden region to write
\begin{equation}
\psi _{0}\left( x\right) =\psi _{0}\left( x_{s}\right)
\exp{\left(-\frac{1}{\hbar }%
\int_{x_{s}}^{x}p_{\mathrm{aux}}\left( y\right) dy\right)}.
\label{vacu3}
\end{equation}

On the other hand, $\psi _{E}\left( x\right) $ is given by Eq. (\ref{eq89})
in Appendix A. If $E$ is close to $E_{0}$, then Eq. (\ref{norm11}) applies,
and we may write
\begin{equation}
\psi _{E}\left( x\right) \sim 2K_{E}\left[ \frac{\tau }{\hbar }\left(
E-E_{0}\right) F_{+}\left( x_{s},x_{R}\right) \exp{\left(\frac{1}{\hbar }%
\int_{x_{s}}^{x}p\left( y\right) dy\right)}+\frac{1}{2}F_{-}\left(
x_{s},x_{R}\right) \exp{\left(-\frac{1}{\hbar }\int_{x_{s}}^{x}p\left( y\right)
dy\right)}\right] .
\end{equation}
Substituting the two previous expressions into the right hand side of Eq. (%
\ref{new1}) we see that we have to compute the two following integrals,
\begin{equation}
J_{\pm }=\int_{x_{s}}^{\infty }dx\;\left[ U_{\mathrm{aux}}\left( x\right)
-U\left( x\right) \right] \exp{\left(-\frac{1}{\hbar }\int_{x_{s}}^{x}\left[ p_{%
\mathrm{aux}}\left( y\right) \pm p\left( y\right) \right] dy\right)}.
\label{vacu3a}
\end{equation}
The integral, $J_{-}$, is dominated by the region near the lower limit,
where $p_{\mathrm{aux}}\left( x\right) $ is close to $p\left( x\right) $ and
we can write
\[
p_{\mathrm{aux}}\left( x\right) -p\left( x\right) \sim \frac{ p_{%
\mathrm{aux}}^{2}\left( x\right) -p^{2}\left( x\right) }{2\sqrt{%
2MU\left( x_{s}\right) }}=\sqrt{\frac{M}{2U\left( x_{s}\right) }}\left[ U_{%
\mathrm{aux}}\left( x\right) -U\left( x\right) +E-E_{0}\right] ,
\]
{}from where we obtain
\begin{equation}
J_{-}=\hbar \sqrt{\frac{2U\left( x_{s}\right) }{M}}-\left( E-E_{0}\right)
\int_{x_{s}}^{\infty }dx\;\exp{\left(-\frac{1}{\hbar }\int_{x_{s}}^{x}\left[ p_{%
\mathrm{aux}}\left( y\right) -p\left( y\right) \right] dy\right)},  \label{vacu3c}
\end{equation}
where the remaining integral is made negligible by an appropriate choice of $%
U_{\mathrm{aux}}$. For the other integral, $J_{+}$, we see that the
corresponding exponential factor in Eq. (\ref{vacu3a}) decays faster than
the exponential factor of $J_{-}$, so that the region which effectively
contributes to the integral is narrower. Since the pre-exponential factor
vanishes at the lower limit, we find $J_{+}\sim 0$. Finally, putting all
these pieces together into the right hand side of Eq. (\ref{new1}) we get to
leading order,
\[
\left( E-E_{0}\right) \left[ C_{E}+2K_{E}\psi _{0}\left( x_{s}\right) \tau
\sqrt{\frac{2U\left( x_{s}\right) }{M}}F_{+}\left( x_{s},x_{R}\right)
\right] =0,
\]
whose solution, assumed smooth, is
\begin{equation}
C_{E}=-2K_{E}\psi _{0}\left( x_{s}\right) \tau \sqrt{\frac{2U\left(
x_{s}\right) }{M}}F_{+}\left( x_{s},x_{R}\right) .  \label{vacu4}
\end{equation}
We note that $C_{E}$ is independent of the choice of $U_{\mathrm{aux}}$
beyond $x_{s}$, as it should.

Thus, we have found the false vacuum wave function in terms of the
energy eigenfunctions of the original problem. The false ground
state is a superposition of energy eigenstates which are fine
tuned in such a way as to
produce destructive interference outside the potential well. Notice that $%
C_{E}$, because of the factor $K_{E}$ \ in Eq. (\ref{vacu4}),
peaks near the energy of the false ground state, and has a strong
dependence on the energy near this ground state energy. A good
approximation for $C_E$ is given by
\begin{equation}
C_E^2=\frac{\varepsilon}{\pi}\frac{1}{(E-E_+)(E-E_-)}.
\label{vacu4a}
\end{equation}

\subsection{Tunneling from the false vacuum}

Let us now compute the tunneling rate assuming that the particle is
described initially by the false ground state $\psi _{0}$. At time $t$, we
have
\begin{equation}
\psi \left( x,t\right) =\int dE\;e^{-iEt/\hbar }C_{E}\psi _{E}\left(
x\right) ,  \label{tunn1}
\end{equation}
The persistence amplitude is
\begin{equation}
\rho \left( t\right) =\int dx\;\psi _{0}^\ast \left( x\right) \psi \left(
x,t\right) =\int dE\;e^{-iEt/\hbar }C_{E}^{2}.  \label{tunn2}
\end{equation}
With the value of $C_E^2$ given by Eq. (\ref{vacu4a}) $\rho(0)=1$.
To perform the integration we can close the contour of integration in the
complex $E$ plane adding an arc at infinity, whereby we pick up the pole $%
E_- $ in $C_{E}^{2}$, therefore $\rho(t) $ goes like
\begin{equation}
\rho(t)\sim \exp \left[ \frac{-t}{4\tau} \exp\left(-\frac{2}{\hbar}S_{0}
\left( x_{out},x_{R}\right) \right)\right],  \label{tunn3}
\end{equation}
(no prefactor) provided $t$ is not too large.
The tunneling rate for this closed system, $%
\Gamma_{closed}$, may be defined from the persistence probability $%
\rho^2(t)\sim \exp(-\Gamma_{closed}t)$, so that $\Gamma_{closed}= (1/2\tau
)\exp (-2S_{0}/\hbar )$, which agrees with the result of the bounce
solution. Note that if we take the classical lowest energy $E=0$, then $%
x_{R}=x_{L}=0$, $x_{out}=x_{exit}$, and $S_{B}=2S(x_{exit},0)$, but $S_0$
here is the action corresponding to a particle with false vacuum energy $E_0$%
, which differs from zero, consequently it differs from $S_B/2$.
This difference is accounted for by the prefactor $a_q$ in the
instanton result. An equivalent way of deriving this result is to
estimate the integral by a stationary phase approximation.

\section{Wigner function and energy representation}

\label{sec3}

An alternative description of a quantum system is given by the
Wigner function in phase space, which is defined by an integral
transform of the density matrix \cite{Wig32,HilEtAl84}. The Wigner
function for a system described by a wave function $\psi(x)$ is
\begin{equation}
W\left( x,p\right) =\int \frac{dy}{2\pi \hbar }\;e^{ipy/\hbar }\;\psi \left(
x-\frac{y}{2}\right) \psi ^{\ast }\left( x+\frac{y}{2}\right),
\label{wigner1}
\end{equation}
where the sign convention is chosen so that a momentum eigenstate $\psi
_{p_{0}}\left( x\right) \sim e^{ip_{0}x/\hbar }/\sqrt{2\pi \hbar } $ becomes
$W_{p_{0}}\left( x,p\right) =(1/2\pi \hbar) \delta \left( p-p_{0}\right) $.
Moreover, it satisfies
\begin{equation}
\int dp\;W\left( x,p\right) =\;\left| \psi \left( x\right) \right|
^{2},\qquad \int dx\;W\left( x,p\right) =\;\left| \int dx\;\frac{%
e^{-ipx/\hbar }}{\sqrt{2\pi \hbar }}\psi \left( x\right) \right| ^{2},
\label{wigner2}
\end{equation}
and it is normalized so that $\int \int dx\;dp\;W(x,p)=1$. Thus
the Wigner function is similar in some ways to a distribution
function in phase space, it is real but, unlike a true
distribution function, it is not positive defined; this is a
feature connected to the quantum nature of the system it
describes.

The Schr\"{o}dinger equation for the wave function $\psi $,
\begin{equation}
\frac{-\hbar ^{2}}{2M}\frac{\partial ^{2}}{\partial x^{2}}\psi +U\left(
x\right) \psi =i\hbar \frac{\partial }{\partial t}\psi ,  \label{wigner3}
\end{equation}
translates into a dynamical equation for the Wigner function, which is
easily derived. In fact, by taking the time derivative of (\ref{wigner1}),
using the Schr\"{o}dinger equation (\ref{wigner3}), and integrating by parts
we have
\begin{eqnarray}
\frac{\partial }{\partial t}W\left( x,p\right)  &=&-\frac{i}{\hbar }\int
\frac{dy}{2\pi \hbar }\;e^{ipy/\hbar }\;\left\{ \left( \frac{-i\hbar p}{M}%
\right) \frac{\partial }{\partial x}\left[ \psi \left( x-\frac{y}{2}\right)
\psi ^{*}\left( x+\frac{y}{2}\right) \right] \right.   \nonumber \\
&&\left. +\psi \left( x-\frac{y}{2}\right) \left[ U\left( x-\frac{y}{2}%
\right) -U\left( x+\frac{y}{2}\right) \right] \psi ^{*}\left( x+\frac{y}{2}%
\right) \right\} .  \nonumber
\end{eqnarray}
For the cubic potential (\ref{potential}) we have $U\left( x-y/2\right)
-U\left( x+y/2\right) =-M\Omega _{0}^{2}xy+(\lambda /2)x^{2}y+(\lambda
/24)y^{3}$ and, noting that $ye^{ipy/\hbar }=-i\hbar \partial
_{p}e^{ipy/\hbar }$ and $y^{3}e^{ipy/\hbar }=i\hbar ^{3}\partial
_{p}^{3}e^{ipy/\hbar }$, we get the equation for the Wigner function
\begin{equation}
\frac{\partial }{\partial t}W\left( x,p\right) =\left[ U^{\prime }\left(
x\right) \frac{\partial }{\partial p}-\frac{p}{M}\frac{\partial }{\partial x}%
+\frac{\lambda }{24}\hbar ^{2}\frac{\partial ^{3}}{\partial p^{3}}\right]
W\left( x,p\right) ,  \label{wigner4}
\end{equation}
which may be interpreted as a quantum transport equation. The first two
terms on the right hand side are just the classical Liouville terms for a
distribution function, the term with the three momentum derivatives is
responsible for the quantum tunneling behavior of the Wigner function in our
problem. A theorem by Pawula \cite{Ris89} states that a transport equation
should have up to second order derivatives at most, or else an infinite
Kramers-Moyal expansion, for non-negative solutions $W(x,p,t)$ to exist. The
above equation for the Wigner function circumvents the implications of the
theorem since it need not be everywhere-positive. Even if we have an
everywhere-positive Gaussian Wigner function at the initial time, the
evolution generated by an equation such as Eq. (\ref{wigner4}) will not keep
it everywhere-positive. Thus, here we see the essential role played by the
non-positivity of the Wigner function in a genuinely quantum aspect such as
tunneling.

\subsection{The energy representation}

Given that a wave function $\psi $ can be represented in terms of the energy
eigenfunctions $\psi_E$, defined by Eq. (\ref{wkb1}), as
\begin{equation}
\psi \left( x\right) =\int dE\;C_{E}\psi _{E}\left( x\right),  \label{ener1}
\end{equation}
we can introduce a corresponding representation for $W(x,p)$ in terms of a
base of functions $W_{E_{1}E_{2}}(x,p)$ in phase space defined by
\begin{equation}
W_{E_{1}E_{2}}\left( x,p\right) =\int \frac{dy}{2\pi \hbar }\;e^{ipy/\hbar
}\;\psi _{E_{1}}\left( x-\frac{y}{2}\right) \psi_{E_{2}}^\ast \left( x+\frac{%
y}{2} \right).  \label{ener3}
\end{equation}
Then $W\left( x,p\right)$ can be written as
\begin{equation}
W\left( x,p\right) =\int dE_{1}dE_{2}\;C_{E_{1}E_{2}}W_{E_{1}E_{2}}\left(
x,p\right),  \label{ener2}
\end{equation}
where, in this case, we have $C_{E_{1}E_{2}}=C_{E_{1}}C_{E_{2}}^{\ast }$. On
the other hand from the definition of $W_{E_{1}E_{2}}(x,p)$ we can write
\[
\int \frac{dxdp}{\hbar }W_{E_{1}E_{2}}^{*}\left( x,p\right) W_{E_{1}^{\prime
}E_{2}^{\prime }}\left( x,p\right) =\int \frac{dxdy}{2\pi \hbar ^{2}}\left\{
\psi _{E_{1}}\left( x-\frac{y}{2}\right) \psi _{E_{2}}\left( x+\frac{y}{2}%
\right) \psi _{E_{1}^{\prime }}\left( x-\frac{y}{2}\right) \psi
_{E_{2}^{\prime }}\left( x+\frac{y}{2}\right) \right\},
\]
where the $p$ integration has been performed. If we now call $z=x-y/2$, $%
z^{\prime }=x+y/2$; then $dxdy=dzdz^{\prime }$, and
\begin{equation}
\int \frac{dxdp}{\hbar }W_{E_{1}E_{2}}^{\ast }\left( x,p\right)
W_{E_{1}^{\prime }E_{2}^{\prime }}\left( x,p\right) =\frac{1}{2\pi \hbar ^{2}%
}\delta \left( E_{1}-E_{1}^{\prime }\right) \delta \left(
E_{2}-E_{2}^{\prime }\right),  \label{ener4}
\end{equation}
which gives the orthogonality properties of the functions $W_{E_1 E_2}$.
This suggests that any Wigner function may be written in this basis as
\begin{equation}
W\left( x,p,t\right) =\int dE_{1}dE_{2}\;C_{E_{1}E_{2}}\left( t\right)
W_{E_{1}E_{2}}\left( x,p\right).  \label{ener5}
\end{equation}

We call this the energy representation of the Wigner function. In this
representation, the master equation or the quantum transport equation (\ref
{wigner4}) is very simple
\begin{equation}
\frac{\partial }{\partial t}C_{E_{1}E_{2}}\left( t\right) =\frac{ -i}{\hbar }
\left( E_{1}-E_{2}\right) C_{E_{1}E_{2}}\left( t\right),  \label{ener6}
\end{equation}
as one can easily verify. One can give an alternative derivation of the
tunneling rate from this equation, by taking the initial condition for the
Wigner function which corresponds to the false vacuum.

\subsection{Tunneling in the energy representation: closed system}
\label{tiercs}

Let us compute here the tunneling rate from the false
vacuum for our closed
quantum system. We assume that our particle at $t=0$ is trapped
into the well of the potential (\ref{potential}) in the false ground state
with the energy $E_{0}$, \textit{i.\;e.}
the ground state of the auxiliary potential
$U_{\mathrm{aux}}$ introduced in Section \ref{vacu}. We know from that
section that the wave function $\psi _{0}$ of this state can be expressed in
terms of the eigenfunctions $\psi _{E}$ by Eq. (\ref{vacu1a}) with the
coefficients $C_{E}$ given by Eq. (\ref{vacu4}), or Eq. (\ref{vacu4a}).
In terms of the reduced
Wigner function, which we may call $W_{0}(x,p)$, this state is easily
described in the energy representation (\ref{ener5}) by the coefficients $%
C_{E_{1}E_{2}}(0)=C_{E_{1}}(0)C_{E_{2}}^{*}(0)$, where $C_{E}(0)$ is
given by Eq. (\ref{vacu4a}). Because the dynamics of the quantum transport
equation is trivial in the energy representation (\ref{ener6}) the time
dependence of the coefficients $C_{E_{1}E_{2}}(t)$ is simply
\begin{equation}
C_{E_{1}E_{2}}\left( t\right) =e^{ -\frac{i}{\hbar}(E_1-E_2)
t} C_{E_{1}E_{2}}\left( 0\right) .  \label{tunnel1}
\end{equation}
Thus,
according to Eq. (\ref{ener5}), the Wigner function at any time is
\begin{equation}
W\left( x,p,t\right) =\int dE_{1}dE_{2}\;e^{
 -\frac{i}{\hbar}(E_1-E_2)t} C_{E_{1}}\left( 0\right) C_{E_{2}}\left(
0\right) W_{E_{1}E_{2}}\left( x,p\right) .  \label{tunnel2}
\end{equation}

{}From this we can compute, in particular, the probability of finding the
particle at the false vacuum at any time. In terms of the false vacuum
Wigner function and the Wigner function of the tunneling system we may
define that probability as
\begin{equation}
\rho^2(t)=2\pi\hbar \int dx\,dp\, W_0(x,p)W(x,p,t).  \label{tunnel2a}
\end{equation}
This equation can be used in a closed as well as in an open system.
For the closed system of Section
\ref{sec2} where the state is described by the wave function $\psi$ of Eq. (%
\ref{tunn1}) and the false vacuum is described by the wave function $\psi_0$
of Eq. (\ref{vacu1a}), the square of the persistence amplitude (\ref{tunn2})
is given, in fact, by Eq. (\ref{tunnel2a}) when the definition of the Wigner
function, \textit{i.\;e.} Eq. (\ref{wigner1}), is used. For the open system
the quantum state is not described by a pure state and, in general, the
Wigner function $W(x,p,t)$ can be written as $W=\sum_i p_iW_i$ where $p_i$
is the probability of finding the system in the state $\phi_i$ and $W_i$ is
the Wigner function for the state $\phi_i$. The definition (\ref{tunnel2a})
leads in this case to $\rho^2(t)=\sum_i p_i |\langle \psi_0
|\phi_i\rangle|^2 $, which is indeed the probability of finding the system
in the state $\psi_0$.

 When the energy representation
(\ref{ener5}) is used Eq. (\ref{tunnel2a}) becomes
\begin{equation}
\rho ^{2}\left( t\right) =\int dE_{1}dE_{2}\;
e^{-\frac{i}{\hbar}(E_1-E_2) t} C_{E_{1}}^{2}\left( 0\right)
C_{E_{2}}^{2}\left(
0\right).  \label{tunnel3a}
\end{equation}
To compute $\rho ^{2}\left( t\right) $ we shall use the stationary
phase approximation. The idea is that the integration paths for
$E_{1}$ and $E_{2}$ may be deformed simultaneously in such a way
that the integrand comes to be dominated by Gaussian peaks. For
late times it is enough to seek the stationary points of
$\frac{i}{\hbar}(E_1-E_2) .$ In principle, we could include
$C_{E_{1}}^{2}$ and $C_{E_{2}}^{2}$ as fast varying components of
the integrand, but these functions are really fast varying in the
vicinity of $E_{-}$ and $E_{+}$. Thus, the stationary phase
condition leads to $\rho^2\sim \exp(-2t\varepsilon/\hbar)$ in
agreement with the persistence amplitude of Eq. (\ref {tunn3}).
The tunneling rate is $\Gamma_{closed}=2\varepsilon/\hbar$.

It is often convenient to give the tunneling rates in terms
of an equivalent thermal activation problem under a potential barrier
of hight $\varepsilon_s$ with a certain
effective escape temperature $T_{esc}$. This escape temperature is defined
\cite{MarDevCla87} from a given tunneling rate, $\Gamma $, by the equation
\begin{equation}
\Gamma \equiv \frac{1}{2\tau }e^{-\varepsilon _{s}/k_{B}T_{esc}}.
\label{tesc}
\end{equation}
For the closed system, either the WKB approximation or the
instanton method  yield the same $T_{esc}$ with a
barrier penetrability $\Lambda$, defined by
$\Gamma_{closed}=(1/2\tau)\exp(-\Lambda)$ (which gives
$\Lambda=(2/\hbar)S_0$ in the WKB approximation);
see Eq. (\ref{D10}) in Appendix D.

In the following
sections we will use the energy representation of the Wigner function to
compute the tunneling rate in a more complex problem involving coupling to an
environment. The dynamics of the transport equation in the energy
representation is simpler than in the phase space representation
and the initial condition is given in terms of the
coefficients (\ref{vacu4}) which we have already computed. The task would be
more difficult starting from the transport equation in phase space, such as
Eq. (\ref{wigner4}), since the third derivative term makes the solution of
the equation very complicated. One has to resort to methods such as those
based on matrix continued fractions in order to compute decay rates from
master equations for open quantum systems with third order derivative terms
\cite{Ris89,VogRis88,RisVog88,GarZue04}.

\section{The open quantum system}

\label{oqs}

Now we assume that our system of interest is coupled to an
environment. As emphasized by Caldeira and Leggett
\cite{CalLeg83b} any quantum macroscopic system can be modelled by
an open quantum system by adjusting the coupling of the system and
environment variables and by choosing appropriate potentials. One
of the main effects of the environment is to induce decoherence
into the system which is a basic ingredient into the quantum to
classical transition \cite
{CalLeg83b,Zur91,PazHabZur93,ZurPaz94,PazZur99,PazZur01}.

The standard way in which the environment is introduced is to
assume that the system is weakly coupled to a continuum set of
harmonic oscillators, with a certain frequency distribution. These
oscillators represent degrees of freedom to which some suitable
variables of the quantum system are coupled. One usually further
assumes that the environment is in thermal equilibrium and that
the whole system-environment is described by the direct product of
the density matrices of the system and the environment at the
initial time, so that there are no initial system-environment
correlations. The macroscopic quantum system is then described by
the reduced density matrix, or equivalently, by the reduced Wigner
function of the open quantum system. This latter function is
defined from the system-environment Wigner function after
integration of the environment variables.

In order to have a working model in a form as simple as possible, but
that captures the main effect of the environment, we will assume
that the reduced Wigner function, which we still call $W(x,p)$,
satisfies the following dynamical
equation,
\begin{equation}
\frac{\partial }{\partial t}W\left( x,p\right) =\left[ U^{\prime }\left(
x\right) \frac{\partial }{\partial p}-\frac{p}{M}\frac{\partial }{\partial x}%
+\frac{\lambda }{24}\hbar ^{2}\frac{\partial ^{3}}{\partial p^{3}}+
\frac{\partial }{\partial p}\left( \gamma p+
\gamma M\sigma ^{2}\frac{\partial }{\partial p}%
+\hbar \Delta\frac{\partial}{\partial x}
\right) \right] W\left( x,p\right) ,  \label{open1}
\end{equation}
where $\gamma $ which has units of inverse time is the dissipation
coefficient, and $\sigma ^{2}$ and $\Delta$ are, respectively, the
normal and anomalous diffusion coefficients. The last three terms
of this equation represent the effect of the environment: the
first describes the dissipation produced into the system and the
other two are the diffusion or noise terms. An interesting limit,
the so-called weak dissipation limit, is obtained when $\gamma
\rightarrow 0$, so that there is no dissipation, but the
coefficient $\gamma \sigma ^{2}$ is kept fixed. We will generally
refer to equation (\ref{open1}) as the quantum Kramers equation,
or alternatively, as the quantum transport equation. This equation
reduces to a classical Fokker-Planck transport equation when
$\hbar =0$: it becomes Kramer's equation \cite{Kra40,Lan69} for a
statistical system coupled to a thermal bath and has the right
stationary solutions.

This equation can be derived \cite
{CalLeg83a,UnrZur89,HuPazZha92,HuPazZha93,HalYu96,CalRouVer03}
assuming the so-called Ohmic distribution for the frequencies of
the harmonic oscillators. In the high temperature limit, $\gamma$
is constant, $\sigma ^{2}=k_{B}T$, and $\Delta\propto
(k_{B}T)^{-1}$, where $k_{B}$ is Boltzmann's constant and $T$ the
bath temperature. Thus at high temperature one can generally
ignore the anomalous diffusion term. In the low temperature limit,
however, the master equation for the reduced Wigner function
involves time dependent dissipation and diffusion coefficients.
Typically the dissipation coefficient $\gamma(t)$ starts with a
zero value and after a short transient time, after which the
system and the environment become correlated, it reaches a
constant value; the normal diffusion coefficient $\sigma^2(t)$
starts also with a zero value it reaches a maximum and after the
short transient time it undergoes a mild oscillatory behavior
until at time scales $t\gg\Omega^{-1}$ reaches a constant positive
asymptotic value; the anomalous diffusion coefficient has a
similar qualitative behavior but its asymptotic large time value
is negative and depends on the cut off frequency. To be specific
\cite{HuPazZha92,ArtEtAl03,LomVil05,AntEtAl05}, at large time
scales the normal diffusion coefficient becomes $\sigma ^{2}\sim
\frac{1}{2}\hbar \Omega _{0}$, and the anomalous diffusion becomes
$\Delta\sim -2\gamma \ln (\Omega_{\mathrm{cut}}/\Omega _{0})$,
where $\Omega_{\mathrm{cut}}$ is a suitable cut off frequency for
the Ohmic environment. Thus, the vacuum fluctuations of the
environment is felt primarily through the anomalous diffusion
coefficient that can have a large magnitude. Note that in a
macroscopic device such as a single Josephson junction biased by a
fixed external current \cite{CalLeg83b,MarDevCla87,Wei93,Tin96}
one assumes an Ohmic environment just to model the junction
resistivity.

Equation (\ref{open1}) is often used to describe the effect of
decoherence produced by the diffusion coefficient to study the
emergence of classical behavior in quantum systems; this is a
topic of recent interest; see Ref. \cite{PazZur01} for a review.
Of particular relevance to our problem is the study of decoherence
in quenched phase transitions \cite{AntLomMon01}, and the effect
of decoherence in quantum tunneling in quantum chaotic systems
\cite{MonPaz00,MonPaz01}, or in a double-well potential
\cite{AntEtAl05}.

The reduced Wigner function $W(x,p)$ describes the quantum state of the open
quantum system, and given a dynamical variable $A(x,p)$ associated to the
system its expectation value in that quantum state is defined by,
\begin{equation}
\left\langle A\left( x,p\right) \right\rangle =\int dxdp\;A\left( x,p\right)
\;W\left( x,p\right) .  \label{open2}
\end{equation}
Then one can easily prove from Eq. (\ref{open1}) that defining,
\begin{equation}
N=\int dxdp\;W\left( x,p\right) ,\qquad \left\langle E\right\rangle =\int
dxdp\;\left( \frac{p^{2}}{2M}+U\left( x\right) \right) W\left( x,p\right) ,
\label{open3a}
\end{equation}
we have $\dot{N}=0$ and $\langle \dot{E}\rangle =-\gamma (\langle
p^{2}/M\rangle -N\sigma ^{2})$. Note that the value of
$\sigma ^{2}\sim \frac{1}{2}\hbar \Omega _{0}$ for the zero temperature case
is reasonable since for the virial theorem the average kinetic energy
is half the energy,
$\left\langle p^{2}/M\right\rangle =N\sigma ^{2}$,
and the averaged energy is conserved $\langle\dot{E}\rangle=0$.

Further insight in the effect of the different terms of Eq.
(\ref{open1}) can be obtained from  the so called linear entropy
$S=1-\mathrm{Tr}\;\rho_{r} ^{2}$, where $\rho_{r} ^{2}$ is the
reduced density matrix of the system. This entropy is also a
measure of decoherence \cite{ZurHabPaz93} since for a pure state
$\rho_{r}^{2}=\rho_{r} $ and $S=0$, whereas for a mixed state
$\mathrm{Tr}\;\rho_{r}^{2}<1$ and $S>0.$ In position
representation  $\rho_{r} =\rho_{r}\left( x,y\right) $ and $
\mathrm{Tr}\;\rho_{r}^{2}=\int dxdy\;\rho_{r} \left( x,y\right)
\rho_{r} \left( y,x\right) $. In terms of the Wigner function, see
Eq. (\ref{wigner1}), $ \rho_{r}\left( x,y\right) =\int dp\;\exp
[-ip( x-y) /\hbar ]W[ (x+y)/2,p]. $ If we now call $X=\left(
x+y\right) /2$ and $u=x-y$, then $dxdy=dXdu$ and we can write
\begin{equation}
\mathrm{Tr}\;\rho_{r}^{2}
=2\pi \hbar \int dXdp\;W^2\left( X,p\right).
\label{open3a1}
\end{equation}
We may now compute $dS/dt$ using Eq. (\ref{open1}). The first three terms
are total derivatives and do not contribute to $dS/dt$. The dissipation term
contributes as
\begin{equation}
\frac{dS_{dis}}{dt}=
-2\pi \hbar \gamma \int dXdp\;W^{2}\left( X,p\right),
\label{open3a2}
\end{equation}
which is negative and may be understood as reduction
of entropy by heat transfer
to the environment. The normal diffusion term
contributes as
\begin{equation}
\frac{dS_{n.dif}}{dt}=4\pi \hbar \gamma M\sigma ^{2}\int dXdp\;\left( \frac{\partial W}{%
\partial p}\right)^{2},
\label{open3a3}
\end{equation}
which is positive and always increase the linear entropy.
The anomalous diffusion term, on the other hand,
contributes as
\begin{equation}
\frac{dS_{a.dif}}{dt}=4\pi \hbar^2\Delta \int dXdp\; \frac{%
\partial W}{\partial p} \frac{\partial W}{\partial X},
\label{open3a4}
\end{equation}
which has no defined sign. We may infer from here that normal diffusion
will always induce decoherence.

\subsection{Energy representation of the reduced Wigner function}

Let us now use the base of functions in phase space $W_{E_{1}E_{2}}(x,p)$,
introduced in Eq. (\ref{ener3}), to represent the reduced Wigner function $%
W\left( x,p,t\right) $ as in Eq. (\ref{ener5}). The previous $N$ and $%
\left\langle E\right\rangle $ have very simple expressions in this
representation:
\begin{equation}
N=\int dE\;C_{EE}\left( t\right) ,\qquad \left\langle E\right\rangle =\int
dE\;E\;C_{EE}\left( t\right) .  \label{open4}
\end{equation}
To check the last equation we note that $\int dxdp\;\left[
(p^{2}/2M)+U\left( x\right) \right] W_{E_{1}E_{2}}\left( x,p\right)
=E_{1}\delta \left( E_{1}-E_{2}\right) ,$ which can be easily proved by
explicit substitution of the definition of $W_{E_{1}E_{2}}$, and trading
powers of $p$ by derivatives with respect to $y$ into expressions
(\ref{ener3}), and partial integrations.

The quantum transport equation (\ref{open1}) in the energy representation
becomes,
\begin{equation}
\frac{\partial }{\partial t}C_{E_{1}E_{2}}\left( t\right) =\frac{ -i }{\hbar
}\left( E_{1}-E_{2}\right) C_{E_{1}E_{2}}\left( t\right) +\int
dE_{1}^{\prime }dE_{2}^{\prime }\;Q_{E_{1}E_{2},E_{1}^{\prime }E_{2}^{\prime
}}C_{E_{1}^{\prime }E_{2}^{\prime }}\left( t\right),  \label{open5}
\end{equation}
where, after one integration by parts,
\begin{equation}
Q_{E_{1}E_{2},E_{1}^{\prime }E_{2}^{\prime }}= -2\pi \hbar ^{2}\int \frac{%
dxdp}{\hbar }\left( \frac{\partial }{\partial p}W_{E_{1}E_{2}}^{*}\left(
x,p\right) \right) \left( \gamma p
+\gamma M\sigma^2\frac{\partial }{\partial p}%
+\hbar\Delta\frac{\partial}{\partial x}
\right) W_{E_{1}^{\prime }E_{2}^{\prime }}\left( x,p\right),  \label{open6}
\end{equation}
which has the contributions from the dissipative,
normal diffusion, and anomalous diffusion
parts, respectively, as
\begin{equation}
Q_{E_1E_2,E_1^{\prime }E_2^{\prime }}= Q_{E_1E_2,E_1^{\prime }E_2^{\prime
}}^{\left( D\right) }+ Q_{E_1E_2,E_1^{\prime }E_2^{\prime }}^{\left(
N\right) }+Q_{E_1E_2,E_1^{\prime }E_2^{\prime }}^{\left(
A\right) }.
\label{open6a}
\end{equation}
{}From Eq. (\ref{ener3}) it is easy to see that these coefficients can all
be written in terms of the following matrix elements:
\begin{eqnarray}
X_{E_1E_2}&=&\int dx\;x\;\psi _{E_1}\left( x\right) \psi _{E_2}\left(
x\right),  \label{open7a} \\
P_{E_1E_2}&=& \frac \hbar i\int dx\;\;\psi _{E_1}\left( x\right) \frac
\partial {\partial x}\psi _{E_2}\left( x\right) ,  \label{open7b} \\
\left( XP\right) _{E_1E_2}&=& \frac \hbar i\int dx\;\;x\psi _{E_1}\left(
x\right) \frac \partial {\partial x}\psi _{E_2}\left( x\right),
\label{open7c} \\
X_{E_1E_2}^2&=&\int dx\;x^2\;\psi _{E_1}\left( x\right) \psi _{E_2}\left(
x\right).  \label{open7d}
\end{eqnarray}
Explicitly, we have that
\begin{equation}
Q_{E_1E_2,E_1^{\prime }E_2^{\prime }}^{\left( D\right) }
=\frac{-i\gamma}{2\hbar }%
\left[ \left( XP\right) _{E_1E_1^{\prime }}\delta \left( E_2-E_2^{\prime
}\right) -P_{E_1E_1^{\prime }}X_{E_2E_2^{\prime }}-X_{E_1E_1^{\prime
}}P_{E_2E_2^{\prime }}+\left( XP\right) _{E_2E_2^{\prime }}\delta \left(
E_1-E_1^{\prime }\right) \right],  \label{open8a}
\end{equation}
\begin{equation}
Q_{E_1E_2,E_1^{\prime }E_2^{\prime }}^{\left( N\right) }
=\frac{\gamma M\sigma^2}{\hbar ^2%
}\left[ 2X_{E_1E_1^{\prime }}X_{E_2E_2^{\prime
}}-X_{E_1E_1^{\prime }}^2\delta \left( E_2-E_2^{\prime }\right)
-X_{E_2E_2^{\prime }}^2\delta \left( E_1-E_1^{\prime }\right)
\right],  \label{open8b}
\end{equation}
\begin{equation}
Q_{E_1E_2,E_1^{\prime }E_2^{\prime }}^{\left( A\right) }
=\frac{\Delta}{\hbar }%
\left[ \left( XP\right) _{E_1E_1^{\prime }}\delta \left(
E_2-E_2^{\prime }\right) -P_{E_1E_1^{\prime }}X_{E_2E_2^{\prime
}}+X_{E_1E_1^{\prime }}P_{E_2E_2^{\prime }}-\left( XP\right)
_{E_2E_2^{\prime }}\delta \left( E_1-E_1^{\prime }\right) \right].
\label{open8c}
\end{equation}
Thus, in terms of the Wigner function elements $C_{E_{1}E_{2}}$
the dynamics of the quantum transport equation (\ref{open5}) is
simple. Note that the coefficients $Q^{\left( D\right) }$ and
$Q^{\left( N\right) }$ preserve parity, while  $Q^{\left( A\right)
}$ and a coefficient $Q^{(Q)}$ (which corresponds to the pure
quantum first term in the equation) change parity; in the sense
that the symmetric and antisymmetric parts of  $C_{E_{1}E_{2}}$
are independently preserved or interchanged when contracted to
these terms.

Equation (\ref{open5}) resembles
a similar equation when a Floquet basis of states are used \cite
{Shi65,MilWya83,BluEtAl91,UteDitHan94}, which are very useful when the
Hamiltonian of the system is periodic in time. The Floquet basis is discrete
in such a case and a numerical evaluation of the corresponding matrix
elements (\ref{open7a})-(\ref{open7d}) can be performed; see for instance
\cite{MonPaz00,MonPaz01} for a recent application. It is remarkable that in
our case approximated analytic expressions for these matrix elements can be
found.

\subsection{Some properties of the matrix elements}

The matrix elements (\ref{open7a})-(\ref{open7d}) have a clear physical
interpretation and several relations can be derived among them. Note that $%
X_{E_1E_2}$ is the matrix element of the position operator $X$ in the energy
representation. Since $X\psi _E\left( x\right) =x\psi _E\left( x\right), $
we must have $\int dE_1\;X_{EE_1}\psi _{E_1}\left( x\right) =x\psi _E\left(
x\right). $

On the other hand, $P_{E_1E_2}$ is the matrix element for the momentum
operator. The canonical commutation relation $\left[ P,X\right] =-i\hbar $,
implies $\left[ H,X\right] =(-i\hbar/M)P$, and taking matrix elements on
both sides we have
\begin{equation}
\left( E_1-E_2\right) X_{E_1E_2}=-\frac{i\hbar }MP_{E_1E_2}.  \label{prop2}
\end{equation}
Also, $X_{E_1E_2}^2$ is the matrix element of $X^2$, therefore
\begin{equation}
X_{E_1E_2}^2=\int dE\;X_{E_1E}X_{EE_2}.  \label{prop3}
\end{equation}
On the other hand, $\left( XP\right) _{E_1E_2}$ is the matrix element of $XP$%
, consequently $\left[ \left( XP\right) _{E_2E_1}\right] ^{*}=-\left(
XP\right) _{E_2E_1}$ corresponds to $PX$, and $\left( XP\right)
_{E_1E_2}+\left( XP\right) _{E_2E_1}=\left[ X,P\right] _{E_1E_2}=i\hbar
\delta \left( E_1-E_2\right) $. Also $\left( XP\right) _{E_1E_2}-\left(
XP\right) _{E_2E_1} =(iM/\hbar) \left( E_1-E_2\right) X_{E_1E_2}^2 $, where
the commutator $[H,X^2 ]$ has been used in the last step, therefore
\begin{equation}
\left( XP\right) _{E_1E_2}=\frac{iM}{2\hbar }\left( E_1-E_2\right)
X_{E_1E_2}^2+\frac{i\hbar }2\delta \left( E_1-E_2\right).  \label{prop4}
\end{equation}
We have, also, that $\left( XP\right) _{E_1E_2}=\int dE\;X_{E_1E}P_{EE_2} $.
One may check, for consistency, that these relations imply $\dot N=0.$ In
Appendix B a test of the quantum transport equation in the energy
representation (and of the above matrix element properties) is given by
checking that a stationary solution with a thermal spectrum is, indeed, a
solution in the high temperature limit.

\subsection{Computing the matrix elements}

\label{computmatrix}

The matrix elements contain singular parts coming from the integrals over
the unbound region beyond $x_{s}.$ These singular parts are easy to compute,
since far enough the wave functions assume the simple form (\ref{norm8}).
When performing the calculation of the singular parts of the matrix elements
we will use that when $\bar x\rightarrow \infty$, we have the identities
\begin{equation}
\frac{\sin (p\bar x/\hbar)}{\pi p}\rightarrow \delta \left( p\right) ,
\qquad \frac{\cos (p\bar x/\hbar)}p\rightarrow 0,  \label{comput1}
\end{equation}
which can be easily checked by taking the Fourier transforms of these
functions with respect to $p$.

The computation of the singular parts of the matrix elements (\ref{open7a})-(%
\ref{open7d}) may be reduced to the evaluation of three basic integrals.
These integrals are
\begin{equation}
A_{A,S}\left( p_{1},p_{2}\right) =\int dx\;\sin \left[ \left( p_{1}\mp
p_{2}\right) x/\hbar+\delta _{1}\mp \delta _{2}\right],  \label{comput2a}
\end{equation}
and
\begin{equation}
B\left( p_{1},p_{2}\right) =\int dx\;\sin \left( p_{1}x/\hbar+\delta
_{1}\right) \sin \left( p_{2}x/\hbar+\delta _{2}\right),  \label{comput2b}
\end{equation}
where, for simplicity, we have written $p_{i}\equiv p_{\infty }\left(
E_{i}\right) $ and $\delta _{i}\equiv\delta \left( E_{i}\right) $ ($i=1,2$).
The matrix element $X_{E_{1}E_{2}}$ is
\begin{eqnarray}
X_{E_{1}E_{2}} &\sim& \frac{2M}{\hbar \pi \sqrt{p_{1}p_{2}}}\int dx\;x\;\sin
\left( p_{1}x/\hbar+\delta _{1}\right) \sin \left( p_{2}x/\hbar+\delta
_{2}\right)  \nonumber \\
\ &=&\frac{M}{\pi \sqrt{p_{1}p_{2}}}\left[ -\frac{\partial A}{\partial p_{1}}%
-\frac{\partial \tilde{A}}{\partial p_{2}}-\left( \frac{\partial \delta _{1}%
}{\partial p_{1}}+\frac{\partial \delta _{2}}{\partial p_{2}}\right)
B\right],  \label{comput3a}
\end{eqnarray}
where $A\equiv (A_S-A_A)/2$ and $\tilde{A}\equiv (A_S+A_A)/2$. The matrix
element $X_{E_{1}E_{2}}^2$ is
\begin{eqnarray}
X_{E_1E_2}^2 &\sim &\frac{2M}{\hbar \pi \sqrt{p_1p_2}}\int dx\;x^2\;\sin
\left( p_1x/\hbar+\delta _1\right) \sin \left( p_2x/\hbar+\delta _2\right)
\nonumber \\
&=&\frac{2M}{\pi \sqrt{p_1p_2}}\left[ -\frac{\partial C}{\partial p_1}%
-\left( \frac{\partial \delta _1}{\partial p_1}\right) \tilde{D}\right],
\label{comput3b}
\end{eqnarray}
where, it is easy to show that $C=(\partial B/\partial p_{1})-(\partial
\delta _{1}/ \partial p_{1}) A$, and that $\tilde{D}=-(\partial A/\partial
p_{1})-(\partial \delta _{1}/ \partial p_{1}) B$. The matrix element $%
P_{E_{1}E_{2}}$ is
\begin{equation}
P_{E_{1}E_{2}} \sim \frac{-iM}{\hbar \pi \sqrt{p_{1}p_{2}}}\left(
p_{1}+p_{2}\right) \tilde{A},  \label{comput3c}
\end{equation}
which according to the relations among matrix elements derived in the
previous subsection is related to $X_{E_{1}E_{2}}$ by Eq. (\ref{prop2}). The
remaining matrix element $(XP)_{E_{1}E_{2}}$, on the other hand, can be
computed from the element $X_{E_{1}E_{2}}^2$ according to Eq. (\ref{prop4})

\subsubsection{The integrals $A(p_{1},p_{2})$ and $B(p_{1},p_{2})$}

Thus, we are finally left with the computation of the integrals (\ref
{comput2a}) and (\ref{comput2b}). The integral $B\left( p_{1},p_{2}\right) $
of Eq. (\ref{comput2b}) is dominated by its upper limit $\bar{x}$
\begin{eqnarray}
B\left( p_{1},p_{2}\right)  &\sim &\frac{1}{2}\int^{\bar{x}}dx\;\cos \left[
\left( p_{1}-p_{2}\right) x/\hbar +\delta _{1}-\delta _{2}\right]   \nonumber
\\
\  &\sim &\frac{1}{2\left( p_{1}-p_{2}\right) }\sin \left[ \left(
p_{1}-p_{2}\right) \bar{x}/\hbar +\delta _{1}-\delta _{2}\right]   \nonumber
\\
\  &\rightarrow &\frac{\pi \hbar }{2}\delta \left( p_{1}-p_{2}\right) ,
\label{comput4a}
\end{eqnarray}
The integrals $A_{A,S}\left( p_{1},p_{2}\right) $ are more subtle. The
integral $A_{S}$ is clearly regular on the diagonal. Since we are interested
mostly on the singular behavior of the matrix elements, we can approximate $%
A_{S}\sim 0.$ On the other hand $A_{A}$ is exactly zero on the diagonal.
Close to the diagonal, the integral is dominated by the region where the
argument of the trigonometric function is small, and thereby the integrand
is non oscillatory. Estimating the upper limit of this region as $\bar{x}%
\sim \hbar \left( p_{1}-p_{2}\right) ^{-1}$, we get
\begin{equation}
A_{A}\sim \hbar ^{-1}\left( p_{1}-p_{2}\right) \bar{x}^{2}+\left( \delta
_{1}-\delta _{2}\right) \bar{x}=\hbar PV\left( \frac{1}{p_{1}-p_{2}}\right)
+...\,,  \label{comput4b}
\end{equation}
where the dots stand for regular terms. Actually, this argument would allow
us to introduce an undetermined coefficient in front of the principal value $%
PV$, but in the next section we show that $\hbar $ is the correct
coefficient, as follows from the canonical commutation relations.

Thus, we are now in the position to give the explicit expressions for the
singular parts of the matrix elements and write, finally, the quantum
transport equation in its explicit form.
This is done in detail in the next section.

\section{The quantum transport equation}

\label{ab}

In this Section we explicitly compute the quantum transport equation (\ref
{open1}) satisfied by the reduced Wigner function in the energy
representation.

\subsection{Matrix elements}

First, we need to compute the matrix elements described in section \ref
{computmatrix}. We begin with the matrix element $X_{E_{1}E_{2}}$ which
according to (\ref{comput3a}) and (\ref{comput4a})-(\ref{comput4b}) can be
written as:
\begin{equation}
X_{E_{1}E_{2}}=\frac{ M\hbar}{\sqrt{p_{1}p_{2}}}\left[ \frac{1}{\pi }\frac{%
\partial }{\partial p_{1}} PV\left(\frac{1}{p_{1}-p_{2}}\right) -\frac{%
\partial \delta _{1}}{\partial p_{1}}\delta \left( p_{1}-p_{2}\right)
+...\right].  \label{ab1}
\end{equation}
We go next to the matrix element $P_{E_{1}E_{2}}$, which from (\ref{comput3c}%
) and (\ref{comput4b}) can be written as,
\begin{equation}
P_{E_{1}E_{2}} =\frac{-iM}{\sqrt{p_{1}p_{2}}}\frac{1}{2\pi }\left(
p_{1}+p_{2}\right) PV\frac{1}{p_{1}-p_{2}}.  \label{ab2}
\end{equation}
These two operators $X$ and $P$ are connected through Eq. (\ref{prop2}). It
is easy to check that the two previous results satisfy this relation. Just
notice that from Eq. (\ref{norm2}) we can write $E_1-E_2=(p_1^2-p_2^2)/2M$
which together with Eq. (\ref{ab1}) for $X_{E_1 E_2}$ lead to $-i\hbar/M$
times the right hand side of Eq. (\ref{ab2}), that is
\[
\left( E_{1}-E_{2}\right) X_{E_{1}E_{2}}=-\frac{i\hbar }{M}P_{E_{1}E_{2}}.
\]
Another check of the previous results is the consistency with the canonical
commutation relations
\begin{equation}
\int dE\;\left(P_{E_1E}X_{EE_2}- X_{E_1E}P_{EE_2} \right) =-i\hbar \delta
\left( E_1-E_2\right) .  \label{ab3}
\end{equation}
This check requires a little more work. First it is convenient to change to
momentum variables and write, $\delta \left( E_{1}-E_{2}\right) =(M/\sqrt{%
p_{1}p_{2}})\delta \left( p_{1}-p_{2}\right). $ Then one needs to compute
the integral
\begin{equation}
I \equiv \hbar \int_{-\infty }^{\infty }dp\; PV\left(\frac{1}{p_{1}-p}%
\right) PV\left( \frac{1}{p-p_{2}}\right)=-\hbar \pi ^{2}\delta \left(
p_{1}-p_{2}\right),  \label{ab4}
\end{equation}
The evaluation of this integral is easily performed using the following
representation of the principal value
\[
PV\left( \frac{1}{p}\right) =\int \frac{d\xi }{2\pi\hbar }\;e^{ip\xi/\hbar
}\left( -i\pi \;\mathrm{sign}\left[ \xi \right] \right),
\]
which is easily proved by taking the Fourier transform of $PV (1/p)$. With
the result of Eq. (\ref{ab4}) it is straightforward to check that the
commutation relation (\ref{ab3}) is an identity within our approximation.
This consistency check is important because it can be used to fix to $\hbar$
the coefficient in front of the principal value of $A_A$ in the argument
leading to Eq. (\ref{comput4b}).

We can now move to the matrix elements for $X^2$. Having an expression for $%
X_{E_{1}E_{2}}$ in Eq. (\ref{ab1}) it is best to compute $X_{E_{1}E_{2}}^{2}
$ directly from the relation (\ref{prop3}) which leads to
\begin{eqnarray}
X_{E_{1}E_{2}}^{2} &=&\frac{M\hbar^2}{\sqrt{p_{1}p_{2}}}\left[ \frac{%
\partial ^{2}}{\partial p_{1}\partial p_{2}}\delta \left( p_{1}-p_{2}\right)
+\frac{1}{\pi }\left( \frac{\partial \delta _{1}}{\partial p_{1}}+\frac{%
\partial \delta _{2}}{\partial p_{2}}\right) \frac{\partial }{\partial p_{2}}
PV\left(\frac{1}{p_{1}-p_{2}}\right) \right.  \nonumber \\
&&\left. +\left( \frac{\partial \delta _{1}}{\partial p_{1}}\right)
^{2}\delta \left( p_{1}-p_{2}\right) +...\right],  \label{ab5}
\end{eqnarray}
where we have used the result (\ref{ab4}) and performed the $E$ integration
or, more precisely, the $p$ integration.

The matrix element $(XP)_{E_{1}E_{2}}=\int dE X_{E_{1}E} P_{EE_{2}}$ can be
analogously obtained from the expressions (\ref{ab1}) and (\ref{ab2}). The
result is
\begin{equation}
\left( XP\right) _{E_{1}E_{2}}= \frac{iM\hbar}{2 \sqrt{p_{1}p_{2}}}\left[
2p_{2}\frac{\partial }{\partial p_{1}}\delta \left( p_{1}-p_{2}\right) +%
\frac{1}{\pi }\frac{\partial \delta _{1}}{\partial p_{1}}\left(
p_{1}+p_{2}\right) PV\left(\frac{1}{p_{1}-p_{2}}\right) +...\right].
\label{ab6}
\end{equation}
A further consistency check of these expressions comes from the property (%
\ref{prop4}), which is satisfied within our approximation.

\subsection{The quantum transport equation and time scales}
\label{abaa}

Finally, we can write the quantum transport equation (\ref{open5}),
in a more explicit form. The
coefficient $Q$ is given by (\ref{open6a}), with
the values of the dissipation and diffusion parts given  by
(\ref{open8a}), (\ref{open8b}) and (\ref{open8c}),
which can be directly computed using the
matrix elements obtained in the previous subsection. It is convenient to
introduce new Wigner function coefficients,
\begin{equation}
C_{E_1E_2}\left( t\right) =\frac {M}{\sqrt{p_1p_2}}C_{p_1p_2}\left( t\right),
\label{ab7}
\end{equation}
and the result is the rather cumbersome expression (\ref{ab8})
given in Appendix \ref{aaa3}. As explained there we can get a
local approximation of the quantum transport equation (\ref{ab8}):
\begin{eqnarray}
\frac{\partial C(P,p,t)}{\partial t} &=&\left(\frac{-iPp}{M\hbar }
+\gamma  \frac{\partial }{\partial P}P+ \gamma M\sigma
^{2}\frac{\partial ^{2}}{\partial P^{2}} +i\Delta p\frac{\partial
}{\partial P}\right)C\left(
P,p,t\right)  \nonumber\\
&&-\gamma M\sigma ^{2}\left( \frac{\partial (\delta_1-\delta_2)}{2\partial P}+%
\frac{\partial (\delta_1+\delta_2)}{\partial p}\right)
^{2}C(P,p,t), \label{ab9}
\end{eqnarray}
where the average and difference momentum variables $P$ and $p$,
defined in Eq. (\ref{ab8b}), have been used and where $C(P,p,t)=
C_{p_1p_2}(t)$. {}From the quantum transport equation (\ref{ab8}),
or its local version Eq. (\ref{ab9}), it is easy to discuss the
different time scales of the problem. The first, of course,
involves the dissipation term which includes the dissipation
coefficient $\gamma$, it defines a time scale $\tau_R\sim
\gamma^{-1}$ which is the relaxation time.

But before we go on with the interpretation of the different terms, it is
important to recall the meaning of the Wigner function
coefficients $C_{p_1 p_2}$, or $%
C_{E_1 E_2}$. First, we note that these coefficients are directly
related to the coefficients $C_E$ of the energy eigenfunctions
which make the tunneling state from the false vacuum in the
isolated system, \textit{i.\;e.} when there is no interaction to
the environment. Thus, the coefficients $C_{E_1 E_2}$ describe the
quantum correlations between the energy eigenfunctions that make
the tunneling system. These coefficients are initially separable
$C_{E_1 E_2}(0)= C_{E_1}(0) C_{E_2}^\ast(0)$. In the isolated
closed system its time evolution, as given by Eq. (\ref{ener6}),
is simply $C_{E_1 E_2}(t)=C_{E_1 E_2}(0)\exp[-i(E_1-E_2)t/\hbar]$,
which means that these correlations keep their amplitude in its
dynamical evolution.

This is very different in the open quantum system as a consequence
of the negative local normal diffusion term in Eq. (\ref{ab8})
which depends on the phase shift derivatives, \textit{i.\;e.} the
last term of Eq. (\ref{ab9}), or equivalently the term
(\ref{phase3a}) when it is written in the $p_1$ and $p_2$
variables. This negative defined term has no effect for the
diagonal coefficients, when $E_{1}=E_{2}$, but its effect is very
important for the off diagonal coefficients. In fact, the
amplitude of the off diagonal coefficients exponentially decays in
time. The time scale can be estimated by taking the derivatives of
the phase shifts $\delta _{i}$ ($i=1,2$) near the false vacuum
energy $E_{0}$, which is where the energy wave functions pile up.
Using Eq. (\ref{phase4}) it is easy to see from expression
(\ref{phase3a}) that this time scale is of the order of
\begin{equation}
\tau _{D}\sim \tau _{R}\left( \frac{\lambda _{B}}{l_{D}}\right) ^{2},
\label{ab12}
\end{equation}
where $\tau _{R}$ is the relaxation time, $\lambda _{B}=\hbar /(2\sigma
\sqrt{M})$ is a characteristic de Broglie wavelength (in the high temperature
case when $\sigma ^{2}=k_B T$ it corresponds to the thermal de Broglie
wavelength), and $l_{D}\sim \alpha ^{2}\hbar \sqrt{E_{0}+U_{\infty }}%
/(\varepsilon \sqrt{M})$ is a characteristic length of the problem with $%
\alpha $ a dimensionless parameter that measures the scale of the energy
differences of the off diagonal coefficient, $E_{1}-E_{2}\sim \alpha
\varepsilon $; so it is of order 1 when the energy differences are of order $%
\varepsilon $.
Thus, the last term of equation (\ref{ab9}) destroys the quantum
correlations of the energy eigenfunctions. The time scale $\tau _{D}$ may be
considered as a decoherence time \cite{Zur91}, and thus the effect on
tunneling of this term may be associated to the effect of decoherence.

Another time scale in the problem is, of course, the tunneling time which
according to (\ref{tunn3}) and (\ref{norm13}) is given by $\tau_{\mathrm{tunn%
}}\sim \hbar/\varepsilon$. Its relation to $\tau_D$ is given by $\tau_D\sim
\tau_{\mathrm{tunn}} /(\alpha^4 D)$, where the dimensionless parameter $D$
is
\begin{equation}
D=\gamma \hbar \sigma ^{2}%
\frac{(E_{0}+U_{\infty })}{\varepsilon ^{3}}.  \label{tunnel5}
\end{equation}

It seems clear that when $\tau_D\ll \tau_{\mathrm{tunn}}$ the coefficients
$C_{E_1 E_2}$ become diagonal very fast and the local approximation to the
transport Eq. (\ref{ab9}) is a useful approximation.

\section{Tunneling in the open quantum system}

\label{tunnel}

We can now compute the tunneling rate from the false vacuum for our open
quantum system. Thus, let us assume that our particle at $t=0$ is trapped
into the well of the potential (\ref{potential}) in the false ground state
with the energy $E_{0}$, \textit{i.\;e.}
the ground state of the auxiliary potential
$U_{\mathrm{aux}}$ introduced in Section \ref{vacu}. We know from that
section that the wave function $\psi _{0}$ of this state can be expressed in
terms of the eigenfunctions $\psi _{E}$ by Eq. (\ref{vacu1a}) with the
coefficients $C_{E}$ given by Eq. (\ref{vacu4}).

Under the assumption that the decoherence time is much shorter
than the tunneling time the quantum transport equation (\ref{ab8})
may be simplified. After a typical decoherence time the Wigner
function coefficients $C_{E_1 E_2}$ become diagonal. We will begin
our leading order approximation assuming that these coefficients
are diagonal, which means that decoherence is almost
instantaneous, and then we will correct it in a perturbative way.

\subsection{The Kramers limit}
\label{kramers}

We are interested in the regime where the decoherence time is much
shorter than the tunneling time, or $D \gg 1 $ \textit{i.\;e.}
when the decoherence term (\ref{phase3a}) suppresses the non
diagonal Wigner function coefficients $C_{p_1 p_2}(t)$ and get a
totally decohered Wigner function. In this case using the average
and difference momentum variables $P$ and $p$, see Eqs.
(\ref{ab8b}), we may approximate these coefficients as $C_{p_1
p_2}(t)\equiv C( P,p,t) \sim f( P,t) \delta ( p)$. The quantum
transport equation (\ref{ab8}), or its local approximation Eq.
(\ref{ab9}), reduces then to
\begin{equation}
\frac{\partial f}{\partial t}=\gamma \frac{\partial }{\partial
P}\left( P+M\sigma^2\frac{\partial }{\partial P}\right) f,
\label{diag1}
\end{equation}
which is the Kramers \cite{Kra40} thermal activation equation for
$f(P,t)$. This is a continuity equation for a distribution $f$
with a flux $ \Phi =-\gamma \left( P+M\sigma^2\frac{\partial
}{\partial P}\right) f $. The computation of the tunneling
amplitude has thus been reduced to the computation of the escape
probability of a particle confined to a potential $U(x)$ defined
in Eq. (\ref{potential}) subject to a damping force $\gamma P$ and
white noise with amplitude $\gamma M\sigma^2$; which corresponds
to a temperature $T= \sigma^2/k_B$. The boundary conditions of
Kramers problem are the usual ones. We assume that the particle is
initially trapped at the potential well. Let us define
$P_0=\sqrt{2MU_\infty}$, we may take $U_\infty =0$ in this section
since the asymptotic value of the potential outside the barrier
has no relevance for the classical activation problem. We will ask
that there is no flux entering into the well so that $\Phi(0)=0$
and $\partial f/\partial P(0)=0$. Moreover, at the separatrix when
$P_s=\sqrt{2M\varepsilon_s}$, where $\varepsilon_s$ is the
potential barrier, $f(P_s)=0$. This means that the number of
particles above the separatrix is negligible.

To find the activation rate we seek normal modes $f=f\left(
P\right) e^{-rt}$ and assume that $r$ is very small. If $r=0$ we
have the stationary solution $ f_{0}=\exp
(-P^{2}/2M\sigma^2) $ which satisfies the boundary condition
at $P=0$ but not at $P_{s}.$ We seek a second stationary solution,
for $r=0$, of the form $ F_{0}=f_{0}f_{s} $ then Eq. (\ref{diag1})
reduces to
\begin{equation}
\gamma M\sigma^2\frac{\partial }{\partial P}\left(f_{0}\frac{\partial }{\partial P}%
f_{s}\right)=0,
\label{diag2}
\end{equation}
which leads to $\partial f_s/\partial P=K/f_0$, where the value of
the constant $K$ is irrelevant, and may be chosen as $K=-1$. The
resulting solution $F_0(P)= f_0\int_P^{P_s}dQf_0^{-1}(Q)$
satisfies the boundary condition at $P_s$ but not at $P=0$. Thus,
we now have two stationary solutions of Eq. (\ref{diag1}), $f_0$
and $F_0$, and we may use the variation of constants method to
find a normal mode solution for $r>0$. Let us write
\begin{equation}
f=\alpha \left( P\right) f_{0}+\beta \left( P\right) F_{0},
\label{diag3}
\end{equation}
with the supplementary condition $ \alpha ^{\prime } f_{0}+\beta
^{\prime } F_{0}=0 $. The function $\alpha(P)$ and $\beta(P)$ are
determined by substitution into Eq. (\ref{diag1}) which leads to
the first order differential equations $ \alpha ^{\prime
}=-(r/\gamma M\sigma^2)\left( \alpha F_{0}+\beta
F_{0}^{2}/f_{0}\right) $ and $ \beta ^{\prime }=(r/\gamma
M\sigma^2)\left( \alpha f_{0}+\beta F_{0}\right) $; the
boundary conditions are $\beta \left( 0\right) =\alpha \left(
P_{s}\right) =0.$ For $r$ very small, we may just approximate
$\alpha =\alpha \left( 0\right) $, and $\beta =0$ in the right
hand sides of those equations and we get
\begin{equation}
\alpha \left( P\right) =\alpha \left( 0\right) \left(
1-\frac{r}{\gamma M\sigma^2}\int_{0}^{P}dQ\;F_{0}\left(
Q\right) \right), \label{diag4}
\end{equation}
which leads to the lowest eigenvalue
\begin{equation}
r=\frac{\gamma ME_{0}}{\int_{0}^{P_{s}}dQ\;F_{0}\left( Q\right) }
\sim \frac{\gamma }{\sqrt{%
\pi }}\sqrt{\frac{\varepsilon
_{s}}{\sigma^2}}\;e^{-\varepsilon_s/\sigma^2}.
\label{diag5}
\end{equation}
This is the escape rate of the particle. We may give this rate in
terms of the effective escape temperature defined in Eq.
(\ref{tesc}) by equating $r$ to $(1/2\tau)\exp (-\varepsilon_s/k_B
T_{esc})$, that is
\begin{equation}
T_{esc}=
\frac{\sigma^2}{k_B}\left[1-\frac{\sigma^2}{\varepsilon_s}
\ln\left(
\frac{2\gamma}{\Omega_0}\sqrt{\frac{\pi\varepsilon_s}{\sigma^2}}
\right)\right]^{-1}, \label{diag6}
\end{equation}
where we have used that the dynamical time $\tau$, defined in Eq.
(\ref{norm12}), is $\tau\sim \pi/\Omega_0$. Thus, we have a escape
temperature of the order of $T_{esc}\sim \sigma^2/k_B\leq
\frac{1}{2}\hbar\Omega_0/k_B$, at zero temperature.

\subsection{Effect of anomalous diffusion}
\label{anomalous}

We may now estimate the effect of anomalous diffusion into the
tunneling rate.  We have seen in section \ref{oqs} that the effect
of anomalous diffusion to the linear entropy of the reduced
density matrix of the open quantum system is undefined, unlike the
effect of normal diffusion. It may increase or decrease the
entropy, and hence the decoherence, depending on the product of
gradients of the Wigner function on phase space; see Eq.
(\ref{open3a4}). Let us go back to the quantum transport equation
(\ref{open5}). The term $Q^{\left( N\right) }$ is dominated by the
decoherence term (\ref{phase3a}); for this reason we may
distinguish a fast and a slow dynamics. The fast dynamics
corresponds to the decay of the non diagonal Wigner function
coefficients. The slow dynamics is the diffusion of the Wigner
function coefficients along the diagonal which will be considered
in the next section and may be described by a Fokker-Planck
equation for a classical distribution function.

To evaluate the effect of $Q^{\left( A\right) }$, let us consider
Wigner function coefficients $C_{E_1 E_2}$ of the form
\begin{equation}
C=C^{diag}+C^{odd}, \label{nomal1}
\end{equation}
where $C^{diag}$ is diagonal and $C^{odd}$ is antisymmetric. Here
we do not include symmetric off diagonal coefficients, such as
$C^{even}$, because we assume that decoherence takes them to zero,
even though other terms generate it. According to the parity
properties of the $Q^{(D)}$, $Q^{(N)}$ and $Q^{(A)}$ defined in
Eqs. (\ref{open8a}), (\ref{open8b}) and (\ref{open8c}), we may
write Eq. (\ref{open5}) as a set of coupled equations:
\begin{eqnarray}
&&\frac \partial {\partial t}C^{diag}=\left( Q^{\left( D\right)
}+Q^{\left(
N\right) }\right) C^{diag}+Q^{\left( A\right) }C^{odd},\\
\nonumber &&\frac \partial {\partial t}C^{odd}=\left( Q^{\left(
D\right) }+Q^{\left( N\right) }\right) C^{odd}+Q^{\left( A\right)
}C^{diag} . \label{normal2a}
\end{eqnarray}
Note that the first term of Eq. (\ref{open5}) does not contribute
to this set of equations: $Q^{\left( Q\right) }$ does not
contribute to the first equation because it vanishes on the
diagonal, nor to the second equation because it destroys diagonal
Wigner functions. Note that the symmetric off diagonal
coefficients $C^{even}$ do not couple to the diagonal coefficients
$C^{diag}$ through terms that contain a $\delta(p)$, thus these
terms are second order with respect to the diagonal coefficients
as analyzed in section \ref{kramers}. Because the leading process
is decoherence, the second equation may be approximated by $ \frac
\partial {\partial t}C^{odd}= \frac{-1}{\tau_{D}}
C^{odd}+Q^{\left( A\right) }C^{diag} $ where $\tau_D$ is a typical
decoherence time scale estimated in section \ref{abaa}; see
Eq. (\ref{ab12}). Note that this is analogous to the
relaxation time approximation for the Boltzmann equation. If
$\tau_{D}$ is short enough, $C^{odd}$ simply trails $C^{diag}$ and
$ C^{odd}\sim \tau_{D}Q^{\left( A\right) }C^{diag} $ and we obtain
an autonomous equation for $C^{diag}$
\begin{equation}
\frac \partial {\partial t}C^{diag}=\left[ Q^{\left( D\right)
}+Q^{\left( N\right) }+\tau_{D}\left( Q^{\left( A\right) }\right)
^2\right] C^{diag}. \label{normal3}
\end{equation}

To compute $\left( Q^{\left( A\right) }\right) ^2$ we go back to
Eq. (\ref{open8c}), we permute the operators $X$ and $P$ so that
we get an expression in terms of $P$ and $P^2$ only. The linear
terms in $P$ vanish near the diagonal and then replace the $P$
with a typical momentum scale, such as $\sqrt{M\sigma ^2}$.
Finally we get
\begin{equation}
\left( Q^{\left( A\right) }\right) _{E_1E_2,E_1^{\prime
}E_2^{\prime }}^2\sim \frac{4\Delta ^2M\sigma ^2}{\hbar ^2}\left(
X_{E_1E_1^{\prime }}^2\delta _{E_2E_2^{\prime }}+\delta
_{E_1E_1^{\prime }}X_{E_2E_2^{\prime }}^2-2X_{E_1E_1^{\prime
}}X_{E_2E_2^{\prime }}\right) ,  \label{normal5}
\end{equation}
which is the same operator as $Q^{\left( N\right) }.$ Thus,
equation (\ref{normal3}) leads to the first of the previous set of
coupled equations when the anomalous term is neglected, but with a
modified normal diffusion coefficient. The effect of anomalous
diffusion is then to lower the normal diffusion coefficient
$\sigma ^2$ to $\sigma _{eff}^2,$ where
\begin{equation}
\sigma _{eff}^2=\left( 1-\frac{4\tau_{D}\Delta ^2}\gamma \right)
\sigma ^2 . \label{normal6}
\end{equation}
This translates into a lowering of the effective escape
temperature of Eq. (\ref{diag6}), namely $T_{esc}\sim
\sigma_{eff}^2/k_B$.
Thus the
escape temperature is always lower than $\sigma^2/k_B$ but it
tends to it when the decoherence time $\tau_D$ goes to zero.
Consequently, the overall effect of anomalous diffusion is to
inhibit tunneling when strong decoherence is assumed.

\section{Conclusions}

\label{conclusions}

To conclude, let us briefly summarize our results. Under the
assumption of strong decoherence and using a real time formalism
we have estimated the tunneling rate for an open
quantum system representing a quantum particle, trapped in a local
minimum of a cubic potential, coupled to an environment. The real
time formalism is based on the master equation for the reduced
Wigner function that describes the open quantum system. Our
computational method involves the introduction of an energy
representation of the reduced Wigner function which is based on
the energy eigenfunctions of the isolated system. The master
equation in this representation, Eq. (\ref{open5}), is an equation
for some Wigner function coefficients that describe the quantum
correlations between eigenfunctions of different energies.

In spite of its apparent straightforwardness, the original master
equation (\ref{open1}) contains a term with three derivatives of
the Wigner function, which makes it quite hard to handle either
analytically or numerically. From this point of view the reduction
of Eq. (\ref{open1}) to Eq. (\ref{open5}), which is then
approximated by Eq. (\ref{ab9}), is already a definite step
forward. Also, the basic process of decoherence is hard to discern
from Eq. (\ref{open1}), while it is clearly depicted in Eq.
(\ref{ab9}). A striking feature of the decay process is that in
last analysis it is robust against the details of the potential
barrier. For example, the Kramers activation rate is dominated by
a single feature of the barrier (its height), and similarly the
quantum tunneling rate for the closed system. This simplicity is
lost in Eq. (\ref{open1}) but regained in Eq. (\ref{ab9}). We see
that the details of the potential (such as the height of the
barrier) enter the problem only through the boundary conditions
and the phase shifts in the energy eigenfunctions. In particular,
the strength of decoherence is determined by the phase shifts near
the false vacuum energy, where the simple approximation of Eq.
(\ref{phase4}) applies.

In our problem, where the particle is initially trapped in the
false vacuum, the master equation involves a term,
(\ref{phase3a}), that destroys the quantum correlations of the
eigenfunctions and is, thus, responsible for decoherence. The
strength of this term is characterized by the dimensionless
parameter $D$, defined in Eq. (\ref{tunnel5}), which is directly
proportional to the energy difference between the false and true
vacua. Under the assumption of strong decoherence the pure quantum
channel to tunneling is partially suppressed; since decoherence
destroys the fine tuning among the energy eigenfunctions that
makes tunneling possible in the isolated closed system. Tunneling
then follows an activation-like channel due to the zero point
fluctuations of the quantum environment. This is similar to the
result recently found in Ref. \cite{AntEtAl05} for a double well
potential, which uses a large scale numerical simulation to solve
the master equation.

Thus the picture we have is the following. For the isolated closed
system tunneling from the ground state goes through the usual
quantum channel which the WKB approach or the instanton approach
reproduce, and that we can equally reproduce using the energy
representation of the master equation; see section \ref{tiercs}.
When the system is coupled to an environment the general effect is
felt as dissipation and diffusion, the latter coming as normal and
anomalous diffusion. These terms will produce essentially two
effects. On the one hand they will produce decoherence to the
system which will tend to suppress tunneling as the system becomes
more classical. On the other hand diffusion will also introduce
noise into the system which will induce tunneling by a mechanism
similar to thermal activation. In general all these mechanism act
simultaneously and their effects cannot be disentangled.

When decoherence is very fast quantum tunneling is strongly
suppressed and activation dominates, to leading order. In section
\ref{oqs} we have seen how the diffusion term, the normal
diffusion and anomalous diffusion affect the linear entropy of the
reduced matrix density of the open quantum system. We see clearly
that normal diffusion tends to always increase the entropy and
thus to induce classicalization and decoherence. This is in fact
what the decoherence term (\ref{phase3a}) in the quantum transport
equation does. The fact that activation from normal diffusion
becomes the dominant decay mechanism when decoherence is fast
enough validates the analysis of vacuum decay in cosmology and
quantum field theory given in Refs.
\cite{CalVer99,CalRouVer01a,CalRouVer02}. On the other hand, these
references show how the analysis given here may be improved by a
more realistic description of the bath and the system-environment
interaction.

We should emphasize that although the master equation in the
energy representation, Eq. (\ref{open5}), is much simpler than in
the standard phase space representation, Eq.(\ref{open1}), it may
be difficult to go beyond the present results by analytic means.
It may still be possible to treat perturbatively the symmetric but
non diagonal terms in the Wigner function coefficients. These
terms were neglected in section \ref{anomalous} where we
considered the effect of anomalous diffusion assuming coefficients
of the form (\ref{nomal1}). The quantum coefficient $Q^{(Q)}$,
which induce pure quantum tunneling in the isolated system, will
have an imprint in those terms.

\begin{acknowledgments}

We are grateful to Daniel Arteaga, Bei-Lok Hu, Fernando Lombardo,
Diana Monteoliva, Renaud Parentani, Ray Rivers and Albert Roura
for interesting discussions and suggestions. This work has been
partially supported by the Research Projects MEC FPA2004-04582,
DURSI 2005SGR-00082 and by Fundaci\'{o}n Antorchas. E.~C.\
acknowledges support from Universidad de Buenos Aires, CONICET,
Fundaci\'{o}n Antorchas and ANPCYT through grant 03-05229.

\end{acknowledgments}
\appendix

\section{WKB solution}

\label{aaa}

In this Appendix we solve the WKB problem posed in section \ref{wkb}. The
starting point are Eqs. (\ref{wkb1}), (\ref{wkb2}) and (\ref{wkb3}) with the
cubic potential of Eq. (\ref{potential}), we have to match the WKB solutions
in the different regions across the potential function.

\subsection{Matching from forbidden to allowed regions}

Let $x_{0}$ be a classical turning point $U\left( x_{0}\right) =E,$ and let $%
U^{\prime }\left( x_{0}\right) <0.$ Then to the left of $x_{0}$ we have a
forbidden region, the two corresponding independent WKB solutions of the
Schr\"{o}dinger equation (\ref{wkb1}) are
\begin{equation}
F_{\pm }\left( x_{0},x\right) =\frac{e^{\pm S\left( x_{0},x\right) /\hbar }}{%
\sqrt{2p\left( x\right) /\hbar }},  \label{wkb4}
\end{equation}
whereas to the right of $x_{0}$ the two independent solutions are
\begin{equation}
G_{\pm }\left( x,x_{0}\right) =\frac{e^{\pm iS\left( x,x_{0}\right) /\hbar }%
}{\sqrt{2p\left( x\right) /\hbar }},  \label{wkb5}
\end{equation}
and we wish to find the corresponding matching conditions. For $x\rightarrow
x_{0}^{-},$ we can Taylor expand the potential around $x_{0}$ and write $%
p\left( x\right) =\kappa \left( x_{0}-x\right) ^{1/2}$ and $S\left(
x_{0},x\right) =\frac{2}{3}\kappa \left( x_{0}-x\right) ^{3/2}$, where we
have introduced $\kappa =\sqrt{2M\left| U^{\prime }\left( x_{0}\right)
\right| }$. Similarly for $x\rightarrow x_{0}^{+},$ we have $p\left(
x\right) =\kappa \left( x-x_{0}\right) ^{1/2}$ and $S\left( x,x_{0}\right) =%
\frac{2}{3}\kappa \left( x-x_{0}\right) ^{3/2}.$

If we write $x-x_{0}=e^{i\pi }\left( x_{0}-x\right) $ then $iS\left(
x,x_{0}\right) =S\left( x_{0},x\right) $ and it would seem that simple
analytical continuation yields $G_{+}\left( x,x_{0}\right) \rightarrow
e^{-i\pi /4}F_{+}\left( x_{0},x\right) $. However, this is impossible,
recall that if we define the flux $J= -i (\psi ^{\ast }\partial_x \psi -\psi
\partial_x \psi ^{\ast }) $ then the Schr\"odinger equation implies flux
conservation $\partial_x J=0$. Now $G_{+}\left( x,x_{0}\right) $ has $J=1$
and therefore it cannot turn into $F_{+}\left( x_{0},x\right) ,$ which is
real, and has $J=0$. Thus, we try instead
\begin{equation}
G_{+}\left( x,x_{0}\right) \rightarrow e^{-i\pi /4}F_{+}\left(
x_{0},x\right) +\beta F_{-}\left( x_{0},x\right),
\end{equation}
and imposing flux conservation we obtain $\beta=(1/2)\exp(i\pi /4)$. We
therefore find the matching conditions
\begin{equation}
e^{\mp i\pi /4}F_{+}\left( x_{0},x\right) +\frac{1}{2}e^{\pm i\pi
/4}F_{-}\left( x_{0},x\right) \rightarrow G_{\pm}\left( x,x_{0}\right),
\label{wkb6}
\end{equation}
{}from were we finally obtain, using Eq. (\ref{wkb5}),
\begin{eqnarray}
F_{+}\left( x_{0},x\right) \rightarrow \frac{1}{\sqrt{2p\left(
x\right)/\hbar }} \cos \left( \frac{1}{\hbar}S\left( x,x_{0}\right) +\frac{%
\pi }{4}\right),  \label{wkb7}
\end{eqnarray}
and
\begin{eqnarray}
F_{-}\left( x_{0},x\right) \rightarrow \frac{2}{\sqrt{2p\left(
x\right)/\hbar }} \sin \left( \frac{1}{\hbar}S\left( x,x_{0}\right) +\frac{%
\pi }{4}\right).  \label{wkb8}
\end{eqnarray}

\subsection{Matching from allowed to forbidden regions}

Now consider the case when $U^{\prime }\left( x_{0}\right) >0.$ To the left
of $x_{0},$ we have an allowed region and the solutions are oscillatory $%
G_{\pm }\left( x_{0},x\right) $, to the right of the turning point we have a
forbidden region and the solutions are a linear combination of (\ref{wkb4}).
By exactly the same procedure of the previous section, after imposing flux
conservation across $x_{0}$ we obtain:
\begin{equation}
G_{\mp }\left( x_{0},x\right) \rightarrow e^{\pm i\pi /4}F_{+}\left(
x,x_{0}\right) +\frac{1}{2}e^{\mp i\pi /4}F_{-}\left( x,x_{0}\right) .
\label{wkb9}
\end{equation}
Note from these equations that the solution that matches a decreasing
exponential is
\begin{equation}
\frac{1}{\sqrt{2p\left( x\right) /\hbar }}\sin \left( \frac{1}{\hbar }%
S\left( x_{0},x\right) +\frac{\pi }{4}\right) \rightarrow \frac{1}{2}%
F_{-}\left( x,x_{0}\right) .
\end{equation}

\subsection{WKB solution for $0<E<\varepsilon _{s}$}

We can now put all this together to write the energy eigenfunctions for our
cubic potential (\ref{potential}) for energies in the range $0<E<\varepsilon
_{s}$. There are three classical turning points in this case $%
x_{L}<x_{R}<x_{out}.$ To the left of $x_{L}$ we have a forbidden zone
extending to $-\infty ,$ so we have
\begin{equation}
\psi _{E}\left( x\right) \sim K_{E}F_{-}\left( x_{L},x\right) ;\qquad
x<x_{L},
\end{equation}
where $K_E$ is a normalization constant to be determined latter. To the
right of $x_{L}$ we have from (\ref{wkb8})
\begin{equation}
\psi _{E}\left( x\right)\sim \frac{2K_{E}}{\sqrt{2p\left( x\right)/\hbar }}%
\sin \left( \frac{1}{\hbar}S\left( x,x_{L}\right) +\frac{\pi }{4}\right),
\end{equation}
which after using the definition (\ref{wkb3}) can be rewritten in the region
$x_{L}<x<x_{R}$ as
\begin{equation}
\psi _{E}\left( x\right) \sim K_{E}\left( e^{i\left( S\left(
x_{R},x_{L}\right)/\hbar -\pi /4\right) }G_{-}\left( x_{R},x\right)
+e^{-i\left( S\left( x_{R},x_{L}\right)/\hbar -\pi /4\right) }G_{+}\left(
x_{R},x\right) \right).
\end{equation}
This expression is in the form suitable for extension to the forbidden
region, that is, to the right of $x_{R}$. Thus, by using (\ref{wkb9}) we
have to the right of $x_{R}$
\begin{equation}
\psi _{E}\left( x\right) \sim 2K_{E}\left[ \cos \left( \frac{1}{\hbar }%
S\left( x_{R},x_{L}\right) \right) F_{+}\left( x,x_{R}\right) +\frac{1}{2}%
\sin \left( \frac{1}{\hbar }S\left( x_{R},x_{L}\right) \right) F_{-}\left(
x,x_{R}\right) \right] ,  \label{eq89}
\end{equation}
which can be rewritten again as
\begin{equation}
\psi _{E}\left( x\right) \sim 2K_{E}\left[ \cos\left(\frac{1}{\hbar} S\left(
x_{R},x_{L}\right) \right) e^{S\left( x_{out},x_{R}\right)/\hbar
}F_{-}\left( x_{out},x\right) +\frac{1}{2}\sin\left( \frac{1}{\hbar} S\left(
x_{R},x_{L}\right) \right) e^{-S\left( x_{out},x_{R}\right)/\hbar
}F_{+}\left( x_{out},x\right) \right],
\end{equation}
which is in a form suitable for extension to the right of $x_{out}$:
\begin{eqnarray}
\psi _{E}\left( x\right) &\sim &\frac{2K_{E}}{\sqrt{2p\left( x\right)/\hbar }%
}\left[ 2 \cos \left(\frac{1}{\hbar} S\left( x_{R},x_{L}\right) \right)
e^{S\left( x_{out},x_{R}\right)/\hbar } \sin \left(\frac{1}{\hbar} S\left(
x,x_{out}\right) +\frac{\pi }{4}\right) \right.  \nonumber \\
&&\left. +\frac{1}{2} \sin\left(\frac{1}{\hbar} S\left( x_{R},x_{L}\right)
\right) e^{-S\left( x_{out},x_{R}\right)/\hbar } \cos \left(\frac{1}{\hbar}
S\left( x,x_{out}\right) +\frac{\pi }{4}\right)\right].  \label{wkb10}
\end{eqnarray}

Note that if we impose the Bohr-Sommerfeld quantization rule
\begin{equation}
S\left( x_{R},x_{L}\right) =\frac{\pi}{2}(1+2n)\hbar,  \label{wkb11}
\end{equation}
only the subdominant, exponential decreasing part survives. This would
correspond to the case when the far right region is forbidden and may be
used to define energies for false states trapped into the potential well, in
particular $n=0$ will correspond to the false ground state.

\subsection{Normalization}

All that remains now is the determination of the normalization constant $%
K_{E}$ which can be done from the normalization of the wave functions. The
eigenfunctions are subject to continuous normalization
\begin{equation}
\int dx\;\psi _{E_{1}}\left( x\right) \psi _{E_{2}}\left( x\right) =\delta
\left( E_{1}-E_{2}\right) .  \label{norm1}
\end{equation}
Since the functions themselves are regular, the singular behavior must come
from the upper limit, see for instance \cite{LanLif77}. For large enough $x,$
we have from Eq. (\ref{wkb3})
\begin{equation}
p\rightarrow p_{\infty }=\sqrt{2M\left( E+U_{\infty }\right) }.
\label{norm2}
\end{equation}

Let us write from Eq. (\ref{wkb4}),
\begin{equation}
S\left( x,x_{out}\right) =p_{\infty }\left( x-x_{out}\right)
+\int_{x_{out}}^{x}dx^{\prime }\;\left[ \sqrt{2M\left( E-U\left( x^{\prime
}\right) \right) }-\sqrt{2M\left( E+U_{\infty }\right) }\right] ,
\label{norm2a}
\end{equation}
if this integral converges, we may take the upper limit of integration to $%
\infty ,$ whereby
\begin{equation}
S\left( x,x_{out}\right) =p_{\infty }x+f\left( E\right) ,  \label{norm3}
\end{equation}
where $f(E)$ stands for the second term of (\ref{norm2a}). Then, for $x\gg
x_{out}$, we can write from (\ref{wkb10}) and (\ref{norm3})
\begin{equation}
\psi _{E}\left( x\right) \sim \frac{\sqrt{2\hbar }K_{E}}{\sqrt{p_{\infty }}}%
\left[ A\left( E\right) \sin \left( \frac{p_{\infty }x}{\hbar }\right)
+B\left( E\right) \cos \left( \frac{p_{\infty }x}{\hbar }\right) \right] ,
\label{norm4}
\end{equation}
where $A(E)$ and $B(E)$ are given by
\begin{eqnarray}
A\left( E\right)  &=&2\cos \left( \frac{1}{\hbar }S\left( x_{R},x_{L}\right)
\right) e^{S\left( x_{out},x_{R}\right) /\hbar }\cos \left( f\left( E\right)
+\frac{\pi }{4}\right)   \nonumber \\
&&-\frac{1}{2}\sin \left( \frac{1}{\hbar }S\left( x_{R},x_{L}\right) \right)
e^{-S\left( x_{out},x_{R}\right) /\hbar }\sin \left( f\left( E\right) +\frac{%
\pi }{4}\right) ,  \label{norm5a}
\end{eqnarray}
\begin{eqnarray}
B\left( E\right)  &=&2\cos \left( \frac{1}{\hbar }S\left( x_{R},x_{L}\right)
\right) e^{S\left( x_{out},x_{R}\right) /\hbar }\sin \left( f\left( E\right)
+\frac{\pi }{4}\right)   \nonumber \\
&&+\frac{1}{2}\sin \left( \frac{1}{\hbar }S\left( x_{R},x_{L}\right) \right)
e^{-S\left( x_{out},x_{R}\right) /\hbar }\cos \left( f\left( E\right) +\frac{%
\pi }{4}\right) .  \label{norm5b}
\end{eqnarray}

Substituting Eq. (\ref{norm4}) into (\ref{norm1}), the singular terms in the
normalization integral are
\[
\int dx\;\psi _{E_{1}}\left( x\right) \psi _{E_{2}}\left( x\right) \sim
\hbar^2 \pi \frac{K_{E_{1}}^{2}}{p_{ 1}}\left[ A^{2}\left( E_{1}\right)
+B^{2}\left( E_{1}\right) \right] \left[ \frac{dp_{1}}{dE_{1}}\right]
^{-1}\delta \left( E_{1}-E_{2}\right),
\]
where the delta function comes from the $x$ integration which brings $\delta
(p_{ 1}-p_{ 2})$ and where we have defined $p_{i}\equiv p_{\infty}(E_i)$ ($%
i=1,2$) and changed from momentum to energy variables according to $p_{
i}dp_{ i}=M dE_{i}$; see Eq. (\ref{norm2}). The normalization condition
reduces to $1$ the coefficient of the delta function above
\begin{equation}
\hbar ^{2}\pi \frac{K_{E_{1}}^{2}}{M}\left[ A^{2}\left( E_{1}\right)
+B^{2}\left( E_{1}\right) \right] =1.  \label{norm6}
\end{equation}
This suggests the introduction of the phase $\delta_E$ as follows,
\begin{equation}
K_{E}A\left( E\right) =\sqrt{\frac{M}{\hbar ^{2}\pi }}\cos \delta _{E},
\qquad K_{E}B\left( E\right) =\sqrt{\frac{M}{\hbar ^{2}\pi }}\sin \delta
_{E}.  \label{norm7}
\end{equation}
Thus, the eigenfunction at $x\gg x_{out}$ is Eq. (\ref{norm8}), that is
\[
\psi _{E}\left( x\right) \sim \sqrt{\frac{2M}{\hbar \pi p_{\infty }}}\sin
\left(\frac{ p_{\infty }x}{\hbar}+\delta _{E}\right).
\]

To work out the constant $K_{E}$ in greater detail we note that form Eqs. (%
\ref{norm5a}) and (\ref{norm5b}) we have
\begin{equation}
A^{2}+B^{2}=4 \cos ^{2}\left(\frac{1}{\hbar}S\left( x_{R},x_{L}\right)
\right) e^{2S\left( x_{out},x_{R}\right)/\hbar }+\frac{1}{4} \sin^2\left(%
\frac{1}{\hbar} S\left( x_{R},x_{L}\right) \right) e^{-2S\left(
x_{out},x_{R}\right)/\hbar },  \label{norm9}
\end{equation}
which is non vanishing as long as $E$ is real. However, if we allow for
complex energies, as is typical of unstable states, it may be zero provided
\begin{equation}
\cos ^{2}\left(\frac{1}{\hbar}S\left( x_{R},x_{L}\right)\right) =\frac{-1}{16%
} \sin ^{2}\left(\frac{1}{\hbar}S\left( x_{R},x_{L}\right) \right)
e^{-4S\left( x_{out},x_{R}\right)/\hbar }.  \label{norm10}
\end{equation}
The left hand side is zero whenever the energy satisfies the Bohr-Sommerfeld
condition (\ref{wkb11}).

\section{Thermal spectrum}

\label{aa}

In this Appendix we check that the quantum transport equation (\ref{open5})
admits a stationary solution with a thermal spectrum. This can be seen as a
test on the restrictions satisfied by the matrix elements (\ref{open7a})-(%
\ref{open7d}) with $\sigma^2=k_B T$.

An unnormalized thermal density matrix in the position representation reads,
\begin{equation}
\rho \left( x,x^{\prime }\right) =\int dE\;e^{-\beta E}\psi _E\left(
x\right) \psi _E\left( x^{\prime }\right),  \label{a1}
\end{equation}
where $\beta=(k_B T)^{-1}$ and its associated Wigner function is
\begin{equation}
W_\beta \left( x,p\right) =\int dE\;e^{-\beta E}W_{EE}\left( x,p\right),
\label{a2}
\end{equation}
which in the energy representation in the base $W_{E_1 E_2}$ of Eq. (\ref
{ener5}) corresponds to the coefficients $C_{E_1E_2}=e^{-\beta E_1}\delta
\left( E_1-E_2\right) $. Inserting this into the transport equation we get
\begin{equation}
\int dE\;e^{-\beta E}Q_{E_1E_2,EE}=0,  \label{a3}
\end{equation}
which after using Eqs. (\ref{open6a}), (\ref{open8a}) and (\ref{open8b}) can
be written in operator language as
\begin{eqnarray*}
0 &=&\frac 1{2M}\frac i\hbar \left( XPe^{-\beta H}-e^{-\beta H}PX-
Pe^{-\beta H}X+Xe^{-\beta H}P \right) \\
&&+\frac 1{\beta \hbar ^2}\left( X^2e^{-\beta H}+e^{-\beta H}X^2-2Xe^{-\beta
H}X\right).
\end{eqnarray*}

At the infinite temperature limit, $\beta =0,$ this is
\begin{equation}
0 =\frac 1M\frac i\hbar \left( XP-PX\right) -\frac 1{\hbar ^2}\left(
X^2H+HX^2-2XHX\right).
\end{equation}
The first term is the commutator which gives $-M^{-1}$, and the second term
can be written as $-\frac 1{\hbar ^2}\left[ X,\left[ X,H\right] \right]$,
which using $\left[ H,X\right] =( \hbar/i ) (P/M)$ is easily seen to cancel
the first term.

\section{Quantum transport equation}

\label{aaa3}

Here we write explicitly the quantum transport equation (\ref{open5}) in the
energy representation. The coefficient $Q$ in Eq. (\ref{open5}) is given by (%
\ref{open6a}), and the values of the dissipative, normal diffusion,
and anomalous diffusion parts of this
coefficient are given, respectively, by
Eqs. (\ref{open8a}), (\ref{open8b}) and (\ref{open8c}).
These parts can be directly written using the matrix elements deduced in
Section \ref{ab}. When the coefficients $C_{p_{1}p_{2}}$ defined in Eq. (\ref
{ab7}) are introduced the transport equation becomes,
\begin{eqnarray}
\frac{\partial C_{p_{1}p_{2}}}{\partial t}
&=&\frac{-i}{2M\hbar }%
(p_{1}^{2}-p_{2}^{2})C_{p_{1}p_{2}}
+\gamma M\sigma ^{2}\left( \frac{\partial ^{2}}{\partial p_{1}^{2}}+\frac{%
\partial ^{2}}{\partial p_{2}^{2}}\right) C_{p_{1}p_{2}}
\nonumber\\
&&+\frac{\gamma }{2}\left( \frac{\partial }{%
\partial p_{1}}p_{1}
+\frac{\partial }{\partial p_{2}}p_2\right) C_{p_{1}p_{2}}
+i\Delta\left( p_{1}\frac{\partial }{%
\partial p_{1}}
-p_2\frac{\partial }{\partial p_{2}}\right) C_{p_{1}p_{2}}
\nonumber \\
&&+\frac{\gamma-i2\Delta }{4\pi ^{2}}\frac{\partial }{\partial p_{1}}\int
dp_{1}^{\prime }dp_{2}^{\prime }\;\left( p_{2}+p_{2}^{\prime }\right)
P(p_{2}-p_{2}^{\prime })P(p_{1}-p_{1}^{\prime })C_{p_{1}^{\prime
}p_{2}^{\prime }}  \nonumber \\
&&+\frac{\gamma +i2\Delta}{4\pi ^{2}}\frac{\partial }{\partial p_{2}}\int
dp_{1}^{\prime }dp_{2}^{\prime }\;\left( p_{1}+p_{1}^{\prime }\right)
P(p_{2}-p_{2}^{\prime })P(p_{1}-p_{1}^{\prime })C_{p_{1}^{\prime
}p_{2}^{\prime }}  \nonumber \\
&&+\frac{2\gamma M\sigma ^{2}}{\pi^2}\frac{\partial^2}{\partial
p_1\partial p_2}
\int dp_{1}^{\prime }dp_{2}^{\prime }%
P\left( p_{1}-p_{1}^{\prime }\right) %
P\left( p_{2}-p_{2}^{\prime }\right) %
C_{p_{1}^{\prime }p_{2}^{\prime }}  \nonumber \\
&&-\gamma M\sigma ^{2}\left( \frac{\partial \delta _{1}}{%
\partial p_{1}}-\frac{\partial \delta _{2}}{\partial p_{2}}\right)
^{2} C_{p_{1}p_{2}}  \nonumber \\
&&+\frac{\gamma }{4\pi }\left( \frac{\partial \delta _{1}}{\partial p_{1}}-%
\frac{\partial \delta _{2}}{\partial p_{2}}\right) \int dp\;\left[ \left(
p_{1}+p\right) P(p_{1}-p)C_{pp_{2}}-\left( p_{2}+p\right)
P(p_{2}-p)C_{p_{1}p}\right]   \nonumber \\
&&+\frac{i\Delta}{2\pi }\left( \frac{\partial \delta _{1}}{\partial p_{1}}-%
\frac{\partial \delta _{2}}{\partial p_{2}}\right) \int dp\;\left[ \left(
p_{1}+p\right) P(p_{1}-p)C_{pp_{2}}+\left( p_{2}+p\right)
P(p_{2}-p)C_{p_{1}p}\right]   \nonumber \\
&&+\gamma M\sigma ^{2}\int \frac{dp_{2}^{\prime }}{\pi }\left( \frac{%
\partial \delta _{2}}{\partial p_{2}}+\frac{\partial \delta _{2^{\prime }}}{%
\partial p_{2}^{\prime }}-2\frac{\partial \delta _{1}}{\partial p_{1}}%
\right) \frac{\partial P(p_{2}-p_{2}^{\prime })}{\partial p_{2}}%
C_{p_{1}p_{2}^{\prime }}  \nonumber \\
&&+\gamma M\sigma ^{2}\int \frac{dp_{1}^{\prime }}{\pi }\left( \frac{%
\partial \delta _{1}}{\partial p_{1}}+\frac{\partial \delta _{1^{\prime }}}{%
\partial p_{1}^{\prime }}-2\frac{\partial \delta _{2}}{\partial p_{2}}%
\right) \frac{\partial P(p_{1}-p_{1}^{\prime })}{\partial p_{1}}%
C_{p_{1}^{\prime }p_{2}},  \label{ab8}
\end{eqnarray}
where we have used the shorthand notation $P(x)\equiv PV(1/x)$.

This equation may be considerably simplified by noticing the
effect of the negative defined local term which depends on the
phase shift derivatives:
\begin{equation}
-\gamma M\sigma ^{2}\left( \frac{\partial \delta _{1}}{%
\partial p_{1}}-\frac{\partial \delta _{2}}{\partial p_{2}}\right)
^{2} C_{p_{1}p_{2}}. \label{phase3a} \end{equation} This term has
no effect on the diagonal Wigner function coefficients, when
$p_1=p_2$, but it exponentially reduces the off diagonal coefficients
$C_{p_1 p_2}$ on a time scale of the decoherence time, $\tau_D$,
as discussed in Section \ref {ab}. This suggests the following
local approximation to the transport
equation (\ref{ab8}) whenever $\tau_D\ll \tau_{\mathrm{tunn}}$.

To derive the local approximation it is best to introduce
new average and difference momentum variables,
\begin{equation}
P=\frac{1}{2}(p_1+p_2),\ \ \ \ p=p_1-p_2, \label{ab8b},
\end{equation} then when substituting into Eq. (\ref{ab8}) we
have two typical terms involving the $p_1^{\prime}$ and
$p_2^{\prime}$ integrations:
\[
\int dUdu\left( 2P+U\right) P(U-u)P(U+u)C\left( P+U,p+2u\right),\
 \int
dUdu\left( p+u\right) P(U-u)P(U+u)C\left( P+U,p+2u\right),
\]
where we have used $C(P,p)=C_{p_{1}p_{2}}$, and
$U=P^{\prime}-P$ and $u=\frac{1}{2}(p^{\prime}-p)$. Now we make
the hypothesis that the $P$ dependence is softer than the $p$
dependence, so within these integrals we can approximate $ C\left(
P+U,p+2u\right) \sim C\left( P,p+2u\right) $. Then
using Eq. (\ref{ab4}) we can integrate $ \int
dU\;P(U-u)P(U+u)=\frac{\pi ^{2}}{2}\delta \left( u\right) $ and $
\int dU\;UP(U-u)P(U+u)=0 $. In this way all the terms in Eq.
(\ref{ab8}) which do not depend on the phase shifts $\delta_1$ and
$\delta_2$ become local and considerably simplified.
For the phase shift terms
we may notice that far from the resonance all terms are negligible
and close to the resonance the local term (\ref{phase3a}) is
clearly dominant so it makes sense to keep only this term.
Finally, Eq. (\ref{ab9}) follows as the local approximation of the
quantum transport equation (\ref{ab8}).

\section{Tunneling rates for the closed system}
\label{tunnelingrates}

In this Appendix we review the calculation of the quantum
mechanical tunneling rate for the closed system, that is, ignoring
the interaction with the environment. The quantum tunneling rate
as given by the instanton calculation
\cite{CalLeg83b,GraWei84,MarDevCla87} is
\begin{equation}
\Gamma _{closed}^{(inst)}=\frac{a_{q}}{2\tau }e^{-\Lambda _{0}},  \label{7.7}
\end{equation}
where $\tau =\pi /\Omega _{0}$, $\Lambda _{0}=S_{B}/\hbar
=18\varepsilon
_{s}/(5\varepsilon _{0})\sim 12.\,\allowbreak 376$, and the prefactor $%
a_{q}=(120\pi \Lambda _{0})^{1/2}\sim 68.306$; here we use as an
example the values $\varepsilon _{s}/k_{B} \sim
589.74\;\mathrm{mK}$, and $\varepsilon _{0}/k_{B} \sim
171.55\;\mathrm{mK}$, from a tunneling experiment for a single
Josephson junction between two superconducting electrodes biased
by an external current reported in Ref. \cite{MarDevCla87}.
 With these values,
the escape temperature defined in Eq. (\ref{tesc}) is
\begin{equation}
T_{esc}^{(inst)}= \frac{ \varepsilon _{0}/k_{B} }{3.6-
(\varepsilon _{0}/\varepsilon _{s})\ln  a_{q} }\sim 72.345\;%
\mathrm{mK}.
\end{equation}

It is interesting to check that this result agrees with the result
we obtain when the dissipation is zero. We can use our WKB result
as obtained in Sec. \ref{sec2}, see Eq. (\ref{tunn3}), to write
\begin{equation}
\Gamma _{closed}^{(WKB)}=\frac{1}{2\tau }e^{-\Lambda },  \label{7.9}
\end{equation}
where $\Lambda =(2/\hbar)S_{0}(x_{out},x_{R})$, with $S_{0}$
defined in Eqs. (\ref{wkb2})-(\ref{wkb3}), where the potential
$U(x)$ is given by Eq. (\ref{potential}).

For a cubic potential, the relationship among the energy $E,$ the
frequency $\Omega $ and the action $S\left( x_{R},x_{L}\right) $
is best given in parametric form,
\begin{equation}
E=2\varepsilon _{s}\zeta \left( k\right),\quad \Omega =\Omega
_{0}f\left( k\right),\quad S\left( x_{R},x_{L}\right)
=\frac{\varepsilon _{s}}{\Omega _{0}}F\left( k\right), \label{D4}
\end{equation}
with $0<k<1$, and
\begin{eqnarray}
&&\zeta \left( k\right) =\frac{1}{8}\left\{ 2+3\frac{\left( 1+k^{2}\right) }{%
\left[Q\left( k\right)\right]^{1/2} }
-\frac{\left( 1+k^{2}\right) ^{3}}{\left[
Q\left( k\right) \right] ^{3/2}}\right\},  \label{D7}\\
&&f\left( k\right) =\left\{ \frac{2}{\pi }\left[ 4Q\left( k\right)
\right] ^{1/4}K\left[ k^{2}\right] \right\} ^{-1},\nonumber\\
&&F\left( k\right) =\frac{27}{8}\left[ \frac{4}{Q\left( k\right)
}\right] ^{5/4}\left\{ a\left( k\right) E\left[ k^{2}\right]
-\left( 1-k^{2}\right) b\left( k\right) K\left[ k^{2}\right]
\right\},  \label{D9}
\end{eqnarray}
where $E\left[ k^{2}\right] $ and $K\left[ k^{2}\right] $ are the
complete elliptic integrals, and we have introduced the functions
$ Q\left( k\right) =(1/4)\left( 1+14k^{2}+k^{4}\right),$ $a\left(
k\right) =(16/15)\left( 2-k^{2}\right) ^{2}-(1/5)\left(
1-k^{2}\right) \left( 21-5k^{2}\right)$ and $b\left( k\right)
=(8/15)\left( 2-k^{2}\right) -\left( 1-k^{2}\right).$

The Bohr-Sommerfeld condition Eq. (\ref{wkb11}) for the ground
state ($n=0$), corresponds to the parameter $k_{GS}$ such that $
F\left( k_{GS}\right) =\pi \varepsilon _{0}/\varepsilon _{s}$
which implies  that $k_{GS}\sim 0.1152$. This corresponds to
$\zeta \left( k_{GS}\right) \sim 0.1423$ and $f\left(
k_{GS}\right) \sim 0.9550,$ while the harmonic approximation for
the potential yields $0.1454$ and $1$, respectively.

To compute the barrier penetrability,
$\Lambda=(2/\hbar)S_0(x_{out},x_R)$, we observe that
$S\left( x_{out},x_{R}\right) $ at energy $E$ is equal to
$S\left( x_{R},x_{L}\right) $ at energy $E_{ref}=\varepsilon
_{s}-E.$ The exchange of $E$ by
$E_{ref}$ is equivalent to the exchange of $k$ by $k_{ref},$ where $%
\zeta \left( k_{ref}\right) = 1/2 -\zeta \left( k\right) .$
For
$k_{GS}$ we obtain $k_{ref}\sim 0.2433$ and $F\left(
k_{ref}\right) \sim 2.4073$. Therefore
\begin{equation}
\Lambda =\frac{\varepsilon _{s}}{\varepsilon _{0}}F\left(
k_{ref}\right) \sim 8.459. \label{D10}
\end{equation}
This is to be compared against the instanton exponent $\Lambda
_{0}-\ln  a_{q} \sim 8.152.$ In terms of the escape temperature,
the WKB approximation yields
\begin{equation}
T_{esc}^{(WKB)}=\frac{ \varepsilon _{0}/k_{B} }{F\left(
k_{ref}\right) -(\varepsilon _{0}/\varepsilon _{s})\ln \left(
\Omega _{GS}/\Omega _{0}\right) }\sim 70. 869\;\mathrm{mK},
\end{equation}
which is in good agreement with the instanton result; here we have
again used the previous numerical results for $\varepsilon_s$ and
$\varepsilon_0$. This agreement, of course, should not be
surprising since for a closed system our method reduces to the
standard WKB calculation. The purpose of this exercise is just to
check the consistency of our calculation and to illustrate how the
instanton and WKB methods compare. That the difference between
$\exp (\Lambda _{0})$ and $\exp (\Lambda )$ is accounted for by
the prefactor $a_{q}$ of Eq. (\ref{7.7}) can be seen analytically
by a perturbative calculation.

\end{document}